\documentclass[aps,preprint,showkeys,footnote,10pt]{revtex4}
\usepackage{eurosym}
\usepackage{amsfonts}
\usepackage{amsmath}
\usepackage{amssymb}
\usepackage{graphicx}
\usepackage{subfigure}
\usepackage{txfonts}
\usepackage{makeidx}
\usepackage{array,multirow}

\newtheorem{theorem}{Theorem}
\newtheorem{acknowledgement}[theorem]{Acknowledgment}

\input{tcilatex}
\begin{document}

\title{Higher order topological matter and fractional chiral states}
\author{L. B. Drissi$^{1,2,3,4,\ast }$, S. Lounis$^{2}$, E. H. Saidi$%
^{1,2,4} $}
\affiliation{{\small 1-LPHE, Modeling \& Simulations, Faculty of Science, Mohammed V
University in Rabat, Morocco}}
\altaffiliation{{\footnotesize {$\ast $\textbf{Corresponding author:} Lalla Btissam Drissi,
E-mail addresses: lalla-btissam.drissi@fsr.um5.ac.ma, l.drissi@fz-juelich.de}} }
\affiliation{{\small 2- Peter Gr\"{u}nberg Institut and Institute for Advanced
Simulation, Forschungszentrum J\"{u}lich \& JARA, Germany}}
\affiliation{{\small 3- CPM, Centre of Physics and Mathematics, Faculty of Science,
Mohammed V University in Rabat, Morocco}}
\affiliation{{\small 4- College of Physical and Chemical Sciences, Hassan II Academy of
Sciences and Technology, Km 4, Avenue Mohammed VI, Rabat, Morocco}}

\begin{abstract}
We develop a chiral anomalous fermion hamiltonian proposal to study the
higher order topological (HOT) phase with chiral symmetry $\mathcal{C}$
fractionalized like $\mathcal{C}_{x}\mathcal{C}_{y}\mathcal{C}_{z}$. First,
we solve the $\mathcal{C}$-chiral symmetry constraint for eight band models
and describe those induced by the partial $\mathcal{C}_{i}$'s. Then, we
determine the explicit expression of fractional states characterising HOT
matter and comment on the relationships amongst them and with the standard
Altland-Zirnbauer gapless modes. We also give characteristic properties of
the gapless fractional states and compute their contribution to the
topological index of the chiral model. The findings of this work are shown
to be crucial for investigating and handling high order topological phase.
\end{abstract}

\maketitle

\section{Introduction}

The latest advance in topological phases of matter has provided a new bridge
between condensed matter physics and lattice field theory where discrete
symmetries like CPT and mirror invariance constitute a basic algorithm for
modeling topological properties \cite{1T,2T,3T,4T}. These symmetry-
protected topological (SPT) phases have been interpreted in terms of
anomalies, and it has been shown that a similar picture holds for SPT phases
with fermions \cite{1S,2S,3S,4S}. This development has reveled a new aspect
of gapless Dirac modes in topological phase transition and conducted to
numerous attempts to avoid the costly numerical task. In this regard, the
physics of graphene with gapless states has taught us several lessons on
zero modes; it is shared by Dirac spinors with fermionic states formally
behaving like up/down quarks of four dimensional lattice chromodynamics
(QCD) \cite{Bo,Be,Cr}, where light quarks show an anomalous quantum Hall
effect in the presence of background fields \cite{D}. In this stride, an
index signature formula characterising higher dimensional CPT invariant
quantum topological phases with reflections has been obtained in \cite{nous}
by combining ideas from anomalous 2D physics and Higgs-fermion tri-coupling.
For other properties including links between the transport property of
electronic wave modes and lattices' symmetries as well as other issues, see
\cite{Oz,Ja,Dr,Ji,No,Dri,Ki,Dris,Yu}.\  \  \  \newline
Recently, a higher order topological (HOT) matter has been discovered as a
new family of topological phases \cite{00C1,00C2,B3,00C22}. In diverse
D-dimensions, HOT matter is characterized by gapless boundary states living
in co-dimension d higher than one. These modes are protected by non local
discrete symmetries such as reflections and inversion invariance \cite%
{00C1,00C2,00D45,00D4555}, rotation symmetry \cite{D00111,D0101} and their
combinations like roto-inversion \cite{00RT1,00RT2}. A simple example of HOT
insulators is the square lattice with open boundaries, where the surface and
edges are gapped, but no gap is observed in the corner states \cite%
{D1,D00111}. Another example is the cube, where the gapless excitations are
localized in the corners or crystal edges \cite{00D44,00D46}\textrm{. }To
identify the higher-order topology, several approaches have been proposed,
such as the nested Wilson loop to detect quantized multipole moments \cite%
{00Z0} and the many-body order parameters applicable not only to bulk
multipoles in crystals,\ but also to interacting fermionic and bosonic
systems \cite{00Z00,00Z001,00Z002,00Z34}. Mathematical tools taking
advantage of physical arguments, such as K-homology in the real space, have
also been considered \cite{00Z341}. Based on phonon-assisted space-time
engineering, Floquet second order topological phases were derived from a
static trivial system \cite{tyuu}.\newline
Despite the intense effort deployed so far, capturing the key features of
HOT matter remains a challenging task attracting enormous attention. The
quantized polarization, characterizing the topological boundary states at
the corners, constitutes the bulk topological index \cite{00Z2}. The Berry
phase, quantized in $%
%TCIMACRO{\U{2124} }%
%BeginExpansion
\mathbb{Z}
%EndExpansion
_{Q}$, describes the corner states in the higher-order symmetry-protected
topological phases \cite{00Z1}. In 2020, a Dirac fermion model coupled to an
$O\left( 2\right) $ Higgs doublet was proposed in \textrm{\cite{E1}} to
describe the topological properties of the 2D- Benalcazar- Bernevig- Hughes
(BBH) lattice exhibiting a topological quadrupole phase. Using a direct
approach based on topological mappings, the signature index of N-dimensional
systems with open boundaries, including 2D and 3D, has been developed \cite%
{nous}. To understand the interplay between symmetries and topology, HOT
insulators based on quantized Wannier centers have been investigated \cite%
{00Z11}.\ Missing results on the engineering of domain walls in topological
tri-hinge matter have been completed in literature using link between graph
theory and the geometry hosting HOTI \cite{nouus}\newline
In this paper, \textrm{we study a novel aspect regarding the relationship
between conventional topological matter,} classified the Altland-Zirnbauer
(AZ) table, and 3-Dimensional HOT matter going beyond AZ classification.
This novel property concerns chiral Hamiltonian models ($\mathcal{C}H_{%
\mathbf{k}}\mathcal{C}^{-1}=-H_{\mathbf{k}}$) and relies on the
factorization of the $\mathcal{C}$ chiral symmetry of AZ matter as the
product of three observables like $\mathcal{C}_{x}\mathcal{C}_{y}\mathcal{C}%
_{z}$. This factorisation, which to our knowledge hasn't been studied in
literature before, leads to a fractionalization of chiral gapless state into
chiral sub-states (fractional chiral states) carrying eigen charges $q_{i}$
under the three $\mathcal{C}_{i}$ operators. To show off the general scheme,
we solve chirality constraint in terms of 26 degrees of freedom and, after
some calculation steps, we end up with the result that gapless 3D corner
waves functions are described by fractional chiral states occupying the
vertices of a cube characterised by mass parameters $m_{x}^{\pm },m_{y}^{\pm
},m_{z}^{\pm }$ with pair members $\left( m_{i}^{+},m_{i}^{-}\right) $
related to each other by the $\mathcal{C}_{i}$'s. We also study the
topological index (IndH) of the chiral eight band models and show that is
indeed an integer given by a sum over the basic units $\dprod \nolimits_{i}%
\frac{1}{2}\left[ sgn(m_{i}^{+})-sgn(m_{i}^{-})\right] .$\newline
The organisation of this paper is as follows: \textrm{In section II}, we
introduce the chiral 3D lattice model and study its algebraic structure by
using chiral symmetry $\mathcal{C}$. We solve the constraint $\mathcal{C}H_{%
\mathbf{k}}\mathcal{C}^{-1}=-H_{\mathbf{k}}$ and study the $\mathcal{C}_{x}%
\mathcal{C}_{y}\mathcal{C}_{z}$\ factorisation of $\mathcal{C}$ symmetry as
well as its $\mathcal{C}_{i}$ fractions. \textrm{In section III,} we study
the anatomy of chiral eight band hamiltonians; first we show that it carries
26 degrees of freedom; then we describe useful properties of their building
blocks. \textrm{In section IV}, we study the HOT matter constraints and work
out their solutions. In \textrm{section V}, we build the right and left
handed states and show that corners are inhabited by fractional chiral
states. We also compute the topological index for HOT matter. \textrm{In
section VI}, we give the conclusion and make comments. \textrm{Last section}
is devoted to an appendix where some useful aspects on our topological
chiral model are detailed.

\section{Chiral models in 3D}

Following the Altland-Zirnbauer (AZ) classification \cite{F1,F2},\textrm{\ }%
topological chiral matter is described by matrix Hamiltonians $H_{\mathbf{k}}
$ that anticommute with the chiral symmetry generator $\mathcal{C}$ namely%
\begin{equation}
\mathcal{C}H_{\mathbf{k}}\mathcal{C}^{-1}=-H_{\mathbf{k}}\qquad
\Leftrightarrow \qquad \mathcal{C}H_{\mathbf{k}}+H_{\mathbf{k}}\mathcal{C}=0
\label{c1}
\end{equation}%
In this section, we consider \textrm{cubic lattices} and use results on
Dirac matter as well as techniques from the Clifford algebra of 8$\times $8
gamma matrices $\Gamma _{A}$ to study candidate solutions for model
hamiltonians $H_{\mathbf{k}}$ in 3D momentum space with coordinate $\mathbf{k%
}=\left( k_{x},k_{y},k_{z}\right) $. First, we describe the generic
structure of hamiltonians $H_{\mathbf{k}}$ with eight energy bands that obey
(\ref{c1}); these hamiltonians have 26 degrees of freedom. Then, we give
properties of the $\Gamma _{A}$'s needed in the deal with the eigenvalue
equation $H_{\mathbf{k}}\Psi _{\mathbf{k}}=E_{\mathbf{k}}\Psi _{\mathbf{k}}$%
. Appendix B deals with the minimality of the eight bands model investigated
in what follows.

\subsection{Eight bands model}

A generic Hamiltonian with eight bands is represented by 8$\times $8
hermitian matrix $\boldsymbol{H}_{\mathbf{k}}$ that has 64 degrees of
freedom; one of them namely tr$\boldsymbol{H}_{\mathbf{k}}$ can be dropped
out as it corresponds just to a shift of the energy eigenvalues; it is the
Fermi energy level E$_{F}$ that we take as zero level. The remaining 63 ones
can be exhibited by expanding the traceless $\boldsymbol{H}_{\mathbf{k}}$ as
a linear combination $\sum_{\nu =1}^{63}F_{\nu }T_{\nu }$ made of two block
factors: $F_{\nu }$ and $T_{\nu }$. The $F_{\nu }$ coefficients are
functions of momentum vector $\mathbf{k}$ and of coupling constants $\mathbf{%
\Delta }=\left( \Delta _{1},\Delta _{2},...\right) $ of the model; i.e $%
F_{\nu }\left( \mathbf{k},\mathbf{\Delta }\right) $. The $T_{\nu }$ matrices
generate the 63 directions in the space of traceless 8$\times $8 matrices;
they encode the algebraic structure of $\boldsymbol{H}_{\mathbf{k}}$; for
that we start by studying them with some details and turn after to $F_{\nu }$%
. \newline
First of all, notice that there are different ways to deal with the $T_{\nu }
$ matrix generators; one of them is given by the algebra of su(8) matrices
having seven charge operators as basic observables known as Cartan charge
operators \textrm{\cite{su8}}; but an interesting realisation of these $%
T_{\nu }$'s is that given by using 8$\times $8 Dirac matrices $\Gamma _{A}$
and their normal products \textrm{\cite{gama}} denoted as $\Gamma
_{A_{1}...A_{n}}=\Gamma _{A_{1}}...\Gamma _{A_{n}}$ with labels $A_{i}$
constrained like $1\leq A_{1}<...<A_{n}\leq 6$. \ In this realisation, the
63 generators $T_{\nu }$ are as collected in the following table
\begin{equation}
\begin{tabular}{|l||l|l|l|l|l|l|}
\hline
$T_{\nu }=$ & $\Gamma _{A}$ & $\Gamma _{AB}$ & $\Gamma _{ABC}$ & $\Gamma
_{7}\Gamma _{AB}$ & $\Gamma _{7}\Gamma _{A}$ & $\Gamma _{7}$ \\ \hline
$63=$ & $6$ & $15$ & $20$ & $15$ & $6$ & $1$ \\ \hline
\end{tabular}
\label{63}
\end{equation}%
This repartition means that 6 matrices amongst the 63 ones are given by $%
\Gamma _{A},$ 15 others by $\Gamma _{AB}$ and so on. With the representation
(\ref{63}), several properties of the matrix Hamiltonian and its spectrum
can be learnt from the properties of the six hermitian $\Gamma _{A}$'s; this
means that the problem of building $H_{\mathbf{k}}$'s has been partially
mapped to the $\Gamma _{A}$'s and their products; the properties of the $%
F_{\nu }\left( \mathbf{k},\mathbf{\Delta }\right) $ factors making $H_{%
\mathbf{k}}$ will be discussed later on. For the algebraic structure of $H_{%
\mathbf{k}}$, recall the six gamma matrices satisfy the Clifford algebra $%
\Gamma _{A}\Gamma _{B}+\Gamma _{B}\Gamma _{A}=2\delta _{AB}$; these matrices
carries basic properties on HOT phases of matter; some of these properties
are given in this paper; for example, their hermitian product $-i\Gamma
_{1}\Gamma _{2}\Gamma _{3}\Gamma _{4}\Gamma _{5}\Gamma _{6}$ defines the so
called gamma seven matrix $\Gamma _{7}$ that \textrm{turns} out to play also
a central role in this study. The matrix $\Gamma _{7}$ is taken below as the
chiral operator $\mathcal{C}$ appearing in (\ref{c1}); i.e:%
\begin{equation}
\Gamma _{7}=\mathcal{C}
\end{equation}%
and obeys $\left( \Gamma _{7}\right) ^{2}=I.$ With this choice of $\mathcal{C%
}$, eq(\ref{c1}) reads as $\Gamma _{7}H_{\mathbf{k}}=-H_{\mathbf{k}}\Gamma
_{7}$ and the problem of constructing chiral hamiltonians is then brought to
looking for the set of generators $T_{\nu }$ anticommuting with $\Gamma _{7}$%
. Because of the anticommutation relation of different $\Gamma _{A}$; it
results that the above 63 matrices (\ref{63}) can be splitted into two
blocks: $\left( i\right) $ the subset of $T_{\nu }$ generators containing $%
\Gamma _{7}$, $\Gamma _{AB}$ and $\Gamma _{7}\Gamma _{AB}$; all of these
matrices commute with $\Gamma _{7}$ and then disregarded as they violate (%
\ref{c1}). $\left( ii\right) $ the subset containing $\Gamma _{A}$ and $%
\Gamma _{ABC}$; these matrices are good as they anticommute with $\Gamma _{7}
$. So, chiral Hamiltonians $H_{\mathbf{k}}$ expressed in terms of the gamma
matrices have the form
\begin{equation}
H_{\mathbf{k}}=\sum_{A=1}^{6}F_{A}\Gamma ^{A}+\sum_{C>B>A\geq
1}F_{ABC}\Gamma ^{ABC}  \label{hk}
\end{equation}%
They have at most 26 coefficients; six $F_{A}$'s and twenty $F_{ABC}$'s. In
case where all these coefficients are non vanishing, the hermitian $H_{%
\mathbf{k}}$ has 8 simple energy eigenvalues $E_{1},...,E_{8}$ with the
property $E_{1}+...+E_{8}=0$ due to tr$H_{\mathbf{k}}=0$. This sum on energy
eigenvalues defines a hypersurface in an eight dimensional space with
coordinates $\left( E_{1},...,E_{8}\right) ;$ above this hypersurface; one,
generally speaking, has the four (particle) energy eigenvalues of the
conducting band and, down it, the four (hole) energy eigenvalues of the
valence band.

\subsubsection{The gamma matrices:\emph{\ }$\Gamma _{A}$}

To get more insight into the algebraic structure of $H_{\mathbf{k}}$, let us
fix a hermitian realisation of the gamma matrices $\Gamma _{A}$. Being 8$%
\times $8 matrices, they can be represented by tensor products of three sets
of 2$\times $2 Pauli matrices $\sigma _{a},\tau _{a}$, $\varrho _{a}$ and
the associated 2$\times $2 identity matrices denoted as $\sigma _{0},\tau
_{0}$, $\varrho _{0}$. We have%
\begin{equation}
\begin{tabular}{lllllll}
$\Gamma _{1}$ & $=$ & $\varrho _{0}\otimes \mathbf{\gamma }_{1}$ & \qquad
,\qquad & $\Gamma _{2}$ & $=$ & $\varrho _{0}\otimes \mathbf{\gamma }_{2}$
\\
$\Gamma _{3}$ & $=$ & $\varrho _{0}\otimes \mathbf{\gamma }_{3}$ & \qquad
,\qquad & $\Gamma _{4}$ & $=$ & $\varrho _{0}\otimes \mathbf{\gamma }_{4}$
\\
$\Gamma _{5}$ & $=$ & $\varrho _{2}\otimes \mathbf{\gamma }_{5}$ & \qquad
,\qquad & $\Gamma _{6}$ & $=$ & $\varrho _{1}\otimes \mathbf{\gamma }_{5}$%
\end{tabular}
\label{27}
\end{equation}%
where the four 4$\times $4 Gamma matrices $\mathbf{\gamma }_{\alpha }$ and
the $\mathbf{\gamma }_{5}$ are given by%
\begin{equation}
\mathbf{\gamma }_{i}=\tau _{2}\otimes \sigma _{i}\quad ,\quad \mathbf{\gamma
}_{4}=\tau _{1}\otimes \sigma _{0}\quad ,\quad \mathbf{\gamma }_{5}=-\tau
_{3}\otimes \sigma _{0}
\end{equation}
It is interesting to split the set eq(\ref{27}) of hermitian gamma matrices $%
\Gamma _{A}$ into two subsets according to their reality ($\Gamma _{A}^{\ast
}=\Gamma _{A}$) or pseudo reality ($\Gamma _{A}^{\ast }=-\Gamma _{A}$) with
respect to complex conjugation operator $K=(^{\ast })$. We have
\begin{equation}
\Upsilon _{i}=\mathbf{\Gamma }_{2i-1}\qquad ,\qquad \Lambda _{i}=\mathbf{%
\Gamma }_{2i}  \label{fl}
\end{equation}%
with $\Upsilon _{i}^{\ast }=-\Upsilon _{i}$ and $\Lambda _{a}^{\ast
}=\Lambda _{a}$. These matrices obey the Clifford algebra relations $%
\Upsilon _{i}\Upsilon _{j}+\Upsilon _{j}\Upsilon _{i}=2\delta _{ij}$ and $%
\Lambda _{a}\Lambda _{b}+\Lambda _{b}\Lambda _{a}=2\delta _{ab}$ as well as $%
\Upsilon _{i}\Lambda _{a}=-\Lambda _{a}\Upsilon _{i}$.

\subsubsection{Gamma seven: $\Gamma _{7}$}

Defined by $-i\Gamma _{1}\Gamma _{2}\Gamma _{3}\Gamma _{4}\Gamma _{5}\Gamma
_{6}$, the gamma seven is a diagonal 8$\times $8 matrix that can be
expressed in different manners: \textrm{First} as
\begin{equation}
\Gamma _{7}=-\varrho _{3}\otimes \tau _{3}\otimes \sigma _{0}  \label{g7}
\end{equation}%
with the order $\varrho /\tau /\sigma $; this ordering is not very
important; but just an indication for our explicit calculations. Notice that
(\ref{g7}) involves two kinds of third Pauli matrices namely $\varrho _{3}$
and $\tau _{3}$; but no $\sigma _{3}$; this means that the eigenvalue charge
$q_{_{\mathbf{\Gamma }_{7}}}$ of the operator $\Gamma _{7}$ is given by the
product $-q_{_{\varrho _{{\small 3}}}}q_{_{\tau _{{\small 3}}}}$ where $%
q_{_{\varrho _{{\small 3}}}}$ and $q_{_{\tau _{{\small 3}}}}$ are the
eigenvalues of $\varrho _{3}$ and $\tau _{3}$ \textrm{respectively}; this
relation is useful in the study of the chirality of a wave function obeying $%
\Gamma _{7}\Psi =q_{_{\mathbf{\Gamma }_{7}}}\Psi $; we can easily build
right handed ($q_{_{\mathbf{\Gamma }_{7}}}=1$) states $\Psi _{R}$ and left
handed ($q_{_{\mathbf{\Gamma }_{7}}}=-1$) states $\Psi _{L}$ by thinking of $%
\Psi $ in terms of tensor product $\mathbf{\zeta }\otimes \mathbf{\eta }%
\otimes \mathbf{\xi }$ which, for commodity, we write it simply as $\mathbf{%
\zeta \eta \xi }$. As illustration, \textrm{we give here two examples},
\begin{equation}
\Psi _{R}=\mathbf{\zeta }_{+}\mathbf{\eta }_{-}\mathbf{\xi }\quad ,\quad
\Psi _{L}=\mathbf{\zeta }_{-}\mathbf{\eta }_{-}\mathbf{\xi }
\end{equation}%
The exhibited $\pm $\ charges are associated with the eigenvalues of $%
\varrho _{3}\mathbf{\zeta }_{\pm }=\pm \mathbf{\zeta }_{\pm }$ and $\tau _{3}%
\mathbf{\eta }_{\pm }=\pm \mathbf{\eta }_{\pm }$; the $\mathbf{\xi }$ has an
indefinite charge as $\Gamma _{7}$ doesn't depend on $\sigma _{3}$. In
practice, the $\mathbf{\zeta }_{\pm }$ and $\mathbf{\eta }_{\pm }$ in above
relations correspond to the following notation that will be used in what
follows: A generic two- component spinor $\mathbf{\lambda }$ can be splited
as a sum $\mathbf{\lambda }_{+}+\mathbf{\lambda }_{-}$ with%
\begin{equation}
\mathbf{\lambda }_{+}=\left(
\begin{array}{c}
\lambda _{+} \\
0%
\end{array}%
\right) \quad ,\quad \mathbf{\lambda }_{+}=\left(
\begin{array}{c}
0 \\
\lambda _{-}%
\end{array}%
\right)   \label{pm}
\end{equation}%
Together with the representations given above, there is another way to
represent the chiral operator $\Gamma _{7}$; this representation turns out
to be interesting for studying HOT matter; instead of the tensor product (%
\ref{g7}), one uses just matrix products given by the factorisation of $%
\Gamma _{7}$ as a normal product of three diagonal 8$\times $8 matrices as
follows
\begin{equation}
\Gamma _{7}=\mathcal{C}_{x}\mathcal{C}_{y}\mathcal{C}_{z}  \label{xyz}
\end{equation}%
This factorisation, to which we also refer to as fractionalization (see
Appendix D for more details), relies on the two following features: \textrm{%
First}, the matrix operator $\Gamma _{7}$ given by (\ref{g7}) is a diagonal
matrix; it is made of diagonal Pauli matrices. \textrm{Second}, $\Gamma _{7}$
is a particular matrix amongst eight possible diagonal 8$\times $8 matrices
including identity $I_{8}$, the $\Gamma _{7}$ and 6 more others constructed
here below. From its definition like $i^{3}\Gamma _{1}\Gamma _{2}\Gamma
_{3}\Gamma _{4}\Gamma _{5}\Gamma _{6}$, it follows that $\Gamma _{7}$ can
indeed be factorised as in (\ref{xyz}) where $\mathcal{C}_{x},\mathcal{C}%
_{y},\mathcal{C}_{z}$ are charge matrix operators given by%
\begin{equation}
\mathcal{C}_{x}=i\Gamma _{1}\Gamma _{2}\quad ,\quad \mathcal{C}_{y}=i\Gamma
_{3}\Gamma _{4}\quad ,\quad \mathcal{C}_{z}=i\Gamma _{5}\Gamma _{6}
\label{lf}
\end{equation}%
The other three \textrm{companions} of $\mathcal{C}_{i}$ are given by $%
\mathcal{C}_{i}^{\prime }=\mathcal{C}_{i}\Gamma _{7}$. The above $\mathcal{C}
$- charge operators are hermitian; commute between them and with $\Gamma _{7}
$; and it happens that they play a role as important as that of gamma seven;
especially when looking for the gapless mode of the hamiltonian (\ref{hk})
by solving $H_{\mathbf{k}}\Psi _{\mathbf{k}}=0$. To that purpose, we think
it interesting to collect below those useful properties on the $\mathcal{C}%
_{i}$'s.

\subsection{Observables $\mathcal{C}_{i}$}

The charge operators $\mathcal{C}_{x},\mathcal{C}_{y},\mathcal{C}_{z}$ are
three commuting hermitian 8$\times $8 matrices generating rotations by $\pi $%
-angles in the planes 1-2, 3-4 and 5-6; they obey the property $\left(
\mathcal{C}_{x}\right) ^{2}=\left( \mathcal{C}_{y}\right) ^{2}=\left(
\mathcal{C}_{z}\right) ^{2}=I_{8}$ and so have two eigenvalues $\pm 1$. By
using eq(\ref{27}), we can express these $\mathcal{C}_{i}$'s as follows%
\begin{equation}
\begin{tabular}{lll}
$\mathcal{C}_{x}$ & $=$ & $-\varrho _{0}\otimes \tau _{0}\otimes \sigma _{3}$
\\
$\mathcal{C}_{y}$ & $=$ & $+\varrho _{0}\otimes \tau _{3}\otimes \sigma _{3}$
\\
$\mathcal{C}_{z}$ & $=$ & $+\varrho _{3}\otimes \tau _{0}\otimes \sigma _{0}$%
\end{tabular}
\label{tt}
\end{equation}%
from which we learn their charges $\theta _{x},\theta _{y},\theta _{_{z}}$
in terms of products of the Pauli charges $q_{_{\varrho _{{\small 3}%
}}},q_{_{\tau _{{\small 3}}}},q_{_{\sigma _{{\small 3}}}}$ namely%
\begin{equation}
\theta _{x}=-q_{_{\sigma _{{\small 3}}}}\quad ,\quad \theta _{y}=q_{_{\tau _{%
{\small 3}}}}q_{_{\sigma _{{\small 3}}}}\quad ,\quad \theta
_{z}=q_{_{\varrho _{{\small 3}}}}  \label{3q}
\end{equation}%
Notice that in eight band model, one distinguishes 8 diagonal matrices
namely: the identity $\varrho _{0}\tau _{0}\sigma _{0}$, the three Pauli
charge operators $Q_{i}$ defined as follows%
\begin{equation}
\begin{tabular}{lll}
$Q_{x}$ & $=$ & $\varrho _{3}\otimes \tau _{0}\otimes \sigma _{0}$ \\
$Q_{y}$ & $=$ & $\varrho _{0}\otimes \tau _{3}\otimes \sigma _{0}$ \\
$Q_{z}$ & $=$ & $\varrho _{0}\otimes \tau _{0}\otimes \sigma _{3}$%
\end{tabular}%
\end{equation}%
with respective eigenvalues $q_{_{x}},q_{_{y}},q_{_{z}}$; three others given
by $Q_{xy},Q_{xz},Q_{yz}$ with $Q_{xy}=\varrho _{3}\tau _{3}\sigma _{0}$ and
so on; and finally $Q_{xyz}=\varrho _{3}\tau _{3}\sigma _{3}$. From this
list, we see that the basic charge operators are the Pauli ones since the
others are given by products; for example $Q_{xy}=Q_{x}Q_{y}$ and the chiral
operator $\Gamma _{7}$ is just $-Q_{xy}$. At the operator level, the
relationship between the $\mathcal{C}_{i}$'s and the Pauli $Q_{i}$'s is as
follows
\begin{equation}
\mathcal{C}_{x}=-Q_{z}\quad ,\quad \mathcal{C}_{y}=Q_{y}Q_{z}\quad ,\quad
\mathcal{C}_{z}=Q_{x}\quad ,\quad \Gamma _{7}=-Q_{x}Q_{y}
\end{equation}%
From these equalities, we learn two interesting relations regarding the
calculation of the eigen charge $q_{_{\mathbf{\Gamma }_{7}}}$; these are
\begin{equation}
q_{_{\mathbf{\Gamma }_{7}}}=\theta _{_{x}}\theta _{_{y}}\theta _{_{z}}
\label{7}
\end{equation}%
defining the fractionalisation of chiral charge, and%
\begin{equation}
q_{_{\mathbf{\Gamma }_{7}}}=-q_{_{\varrho _{{\small 3}}}}q_{_{\tau _{{\small %
3}}}}=-q_{_{x}}q_{_{y}}
\end{equation}%
Below, we often use the second one.

\subsubsection{Eigenstates of $\mathcal{C}_{i}$ operators}

We give below the eigenstates $\Psi $ of the operators $\mathcal{C}_{x},%
\mathcal{C}_{y},\mathcal{C}_{z}$ solving the eigenvalue equations $\mathcal{C%
}_{i}\Psi =\theta _{i}\Psi $ with $i=x,y,z$ and $\theta _{i}=\pm 1$.

$\bullet $ \emph{case} $\theta _{i}=+1$%
\begin{equation}
\begin{tabular}{l|l}
$\mathcal{C}_{i}$ & $\Psi $ \\ \hline \hline
$\mathcal{C}_{x}$ & $\left( \mathbf{\zeta }_{+}\mathbf{\eta }_{+}+\mathbf{%
\zeta }_{+}\mathbf{\eta }_{-}+\mathbf{\zeta }_{-}\mathbf{\eta }_{+}+\mathbf{%
\zeta }_{-}\mathbf{\eta }_{-}\right) \mathbf{\xi }_{-}$ \\ \hline
$\mathcal{C}_{y}$ & $\left( \mathbf{\zeta }_{+}+\mathbf{\zeta }_{-}\right)
\left( \mathbf{\eta }_{-}\mathbf{\xi }_{-}+\mathbf{\eta }_{+}\mathbf{\xi }%
_{+}\right) $ \\ \hline
$\mathcal{C}_{z}$ & $\mathbf{\zeta }_{+}\left( \mathbf{\eta }_{+}\mathbf{\xi
}_{+}+\mathbf{\eta }_{+}\mathbf{\xi }_{-}+\mathbf{\eta }_{-}\mathbf{\xi }%
_{+}+\mathbf{\eta }_{-}\mathbf{\xi }_{-}\right) $%
\end{tabular}%
\end{equation}

$\bullet $ \emph{case} $\theta _{i}=-1$%
\begin{equation}
\begin{tabular}{l|l}
$\mathcal{C}_{i}$ & $\Psi $ \\ \hline \hline
$\mathcal{C}_{x}$ & $\left( \mathbf{\zeta }_{+}\mathbf{\eta }_{+}+\mathbf{%
\zeta }_{+}\mathbf{\eta }_{-}+\mathbf{\zeta }_{-}\mathbf{\eta }_{+}+\mathbf{%
\zeta }_{-}\mathbf{\eta }_{-}\right) \mathbf{\xi }_{+}$ \\ \hline
$\mathcal{C}_{y}$ & $\left( \mathbf{\zeta }_{+}+\mathbf{\zeta }_{-}\right)
\left( \mathbf{\eta }_{+}\mathbf{\xi }_{-}+\mathbf{\eta }_{-}\mathbf{\xi }%
_{+}\right) $ \\ \hline
$\mathcal{C}_{z}$ & $\mathbf{\zeta }_{-}\left( \mathbf{\eta }_{+}\mathbf{\xi
}_{+}+\mathbf{\eta }_{+}\mathbf{\xi }_{-}+\mathbf{\eta }_{-}\mathbf{\xi }%
_{+}+\mathbf{\eta }_{-}\mathbf{\xi }_{-}\right) $%
\end{tabular}%
\end{equation}

\subsubsection{Eigenstates of $\Gamma _{7}$}

Since $q_{_{\mathbf{\Gamma }_{7}}}$ can take two values $\pm 1$; we
distinguish two chiral eigenstates namely right handed wave function $\Psi
_{R}$ for $q_{_{\mathbf{\Gamma }_{7}}}=1$, and left handed wave function $%
\Psi _{L}$ for $q_{_{\mathbf{\Gamma }_{7}}}=-1$. By using the relationship $%
q_{_{\mathbf{\Gamma }_{7}}}=-q_{_{x}}q_{_{y}}$, it results the two following:%
\begin{equation}
\Psi _{R}=\left( \mathbf{\zeta }_{+}\mathbf{\eta }_{-}+\mathbf{\zeta }_{-}%
\mathbf{\eta }_{+}\right) \left( \mathbf{\xi }_{+}+\mathbf{\xi }_{-}\right)
\label{pr}
\end{equation}%
and%
\begin{equation}
\Psi _{L}=\left( \mathbf{\zeta }_{+}\mathbf{\eta }_{+}+\mathbf{\zeta }_{-}%
\mathbf{\eta }_{-}\right) \left( \mathbf{\xi }_{+}+\mathbf{\xi }_{-}\right)
\label{pl}
\end{equation}%
These wave functions are not yet standardized; they still need to be
normalised.

\section{Anatomy of the chiral hamiltonian}

In this section, we study particular properties of the $F_{A}$ and $F_{ABC}$
coefficients of the chiral hamiltonian (\ref{hk}). First, we describe their
building blocks; after that we use time reversing symmetry (TRS) and
particle hole counterpart to study their local expressions.

\subsection{ Building blocks}

By using the splitting (\ref{fl}), the 20 generators $\Gamma _{ABC}$
decompose in 10 pseudo-real terms given by $\Upsilon _{{\small 123}}$, $%
\Upsilon _{i}\Lambda _{ab}$; and 10 real ones namely $\Lambda _{{\small 123}%
},$ $\Upsilon _{ij}\Lambda _{a}$; see table \textbf{\ref{tb}}. The factor $%
\Lambda _{ab}$ is equal to $i\Lambda _{b}\Lambda _{b}$ and its homologue $%
\Upsilon _{{\small jl}}$ reads as $i\Upsilon _{j}\Upsilon _{l}$; both of $%
\Lambda _{ab}$ and $\Upsilon _{{\small jl}}$ are hermitian. Analogously, the
$\Upsilon _{{\small 123}}$ is given by the product $-i\Upsilon _{{\small 1}%
}\Upsilon _{{\small 2}}\Upsilon _{{\small 3}}$ and $\Lambda _{{\small 123}}$
by the product $-i\Lambda _{{\small 1}}\Lambda _{{\small 2}}\Lambda _{%
{\small 3}}$. As for $\Gamma _{ABC}$, one can also decompose the real $%
F_{ABC}$ in a similar manner as collected in the following table
\begin{equation}
\begin{tabular}{|l||l|l||l|l|}
\hline
$\Gamma _{{\small ABC}}$ & $\Upsilon _{{\small 123}}$ & $\Upsilon
_{i}\Lambda _{ab}$ & $\Lambda _{{\small 123}}$ & $\Upsilon _{ij}\Lambda _{a}$
\\ \hline
$F_{{\small ABC}}$ & $f_{{\small 123}}$ & $\ f_{iab}^{\prime }$ & $g_{%
{\small 123}}$ & $\ g_{ija}^{\prime }$ \\ \hline
$\ 20$ & $\  \ 1$ & $\  \ 9$ & $\  \ 1$ & $\  \ 9$ \\ \hline
\end{tabular}
\label{tb}
\end{equation}%
Using (\ref{27}), one can write down the expression of these matrices in
terms of the Pauli ones; for the examples of $\Upsilon _{{\small 123}}$ and $%
\Lambda _{{\small 123}}$, we have
\begin{equation}
\Upsilon _{{\small 123}}=\varrho _{2}\otimes \tau _{3}\otimes \sigma
_{2}\qquad ,\qquad \Lambda _{{\small 123}}=\varrho _{1}\otimes \tau
_{0}\otimes \sigma _{2}
\end{equation}%
Putting the expressions of (\ref{tb}) back into (\ref{hk}), we end up with
the most general chiral model with eight bands%
\begin{equation}
\begin{tabular}{lll}
$H_{\mathbf{k}}$ & $=$ & $\sum_{i=1}^{3}\Upsilon
_{i}f_{i}+\sum_{i=1}^{3}\Upsilon _{i}\left( \sum_{b>a=1}^{3}\Lambda
_{ab}f_{iab}^{\prime }\right) +f_{{\small 123}}\Upsilon _{{\small 123}}+$ \\
&  & $\sum_{a=1}^{3}\Lambda _{a}g_{a}+\sum_{a=1}^{3}\Lambda _{a}\left(
\sum_{j>i=1}^{3}g_{ija}^{\prime }\Upsilon _{ij}\right) +g_{{\small 123}%
}\Lambda _{{\small 123}}$%
\end{tabular}%
\end{equation}%
In the cases where parts of the $F_{ABC}$ coefficients vanish, some of the
energy eigenvalues $E_{1},...,E_{8}$ may have non trivial multiplicities;
the most singular model is the one with the vanishing of all $F_{ABC}=0$; in
this situation we have two energy eigenvalues $E_{\pm }$ with multiplicity
4; from this view the $F_{ABC}$ coefficients can be interpreted as external
fields lifting the energy eigenvalue degeneracies. Indeed, by setting $%
F_{ABC}=0$ in above chiral hamiltonian; one ends up with the reduction $%
\sum_{A=1}^{6}F_{A}\Gamma ^{A}$ having two eigenvalues $E_{\pm }=\pm \frac{1%
}{2}E_{g}$ and a gap energy as $E_{g}=2(F_{1}^{2}+...+F_{6}^{2})^{1/2}$.
This reduced hamiltonian matrix can be cast as $F_{2i-1}\mathbf{\Gamma }%
^{2i-1}+F_{2i}\mathbf{\Gamma }^{2i}$ where odd and even labels are splitted;
by setting $f_{i}=F_{2i-1}$ and $g_{i}=F_{2i}$ and using (\ref{fl}), it
takes the following form%
\begin{equation}
H_{\mathbf{k}}=\sum_{i=1}^{3}f_{i}\Upsilon ^{i}+\sum_{i=1}^{3}g_{i}\Lambda
^{i}  \label{31}
\end{equation}%
where the 3+3 coefficients $f_{i}$ and $g_{i}$\ are functions of 3D momentum
vector $\left( k_{x},k_{y},k_{y}\right) $\ and the coupling parameters of
the eight band model.

\subsection{Mass matrix and gap energy}

We begin by rewriting $H_{\mathbf{k}}$ in an equivalent useful form; then we
describe the effect of the partial chiral symmetries generated by the $%
\mathcal{C}_{i}$'s. By factorising $\Upsilon ^{i}$ in (\ref{31}) and using
\begin{equation}
\Lambda _{1}=-i\Upsilon _{1}\mathcal{C}_{x}\quad ,\quad \Lambda
_{2}=-i\Upsilon _{2}\mathcal{C}_{y}\quad ,\quad \Lambda _{3}=-i\Upsilon _{3}%
\mathcal{C}_{z}
\end{equation}%
the \textrm{Hamiltonian} (\ref{31}) becomes%
\begin{equation}
H_{\mathbf{k}}=\sum_{l=1}^{3}\Upsilon ^{l}\left( f_{l}-i\phi _{l}\right)
\end{equation}%
where we have set
\begin{equation}
\phi _{x}=g_{1}\mathcal{C}_{x}\quad ,\quad \phi _{y}=g_{2}\mathcal{C}%
_{y}\quad ,\quad \phi _{z}=g_{3}\mathcal{C}_{z}
\end{equation}%
This is a remarkable writing in the sense that the three mass terms $\phi
_{l}$ are matrices with eigenvalues given by functions of the $\theta _{l}$
charges of $\mathcal{C}_{l}$. To exhibit further the effect of these $%
\mathcal{C}_{l}$'s we compute the square of (\ref{31}), we find ($%
f_{i}^{2}+g_{i}^{2}$)$I_{8}$ which is non sensitive to the change $\left(
f_{i},g_{i}\right) $ into $\left( -f_{i},-g_{i}\right) $ which is just the
chiral symmetry of the hamiltonian. But this transformation hides an
interesting feature on which we want to shed light; it concerns the partial
change
\begin{equation}
\left( f_{x},g_{x};f_{y},g_{y};f_{z},g_{z}\right) \quad \rightarrow \quad
\left( -f_{x},-g_{x};f_{y},g_{y};f_{z},g_{z}\right)  \label{fg}
\end{equation}%
where only the signs of $f_{x},g_{x}$ have been changed. This transformation
is clearly not \textrm{the full} chiral symmetry which given by the symmetry
group $\mathbb{Z}_{2}^{x}\times \mathbb{Z}_{2}^{y}\times \mathbb{Z}_{2}^{z}$
\textrm{described in} \textrm{Appendix }\textbf{D}. \textrm{It is a
particular sub-symmetry generated by} $\mathcal{C}_{x}$\textrm{\ (
heuristically speaking, 1/3 of chiral symmetry)}. To our knowledge, this
exotic symmetry has not considered in literature before. Similar
transformations hold also for the $\mathcal{C}_{y}$ and $\mathcal{C}_{z}$;
they fill the missing 2/3 in eq(\ref{fg}) \textrm{as shown on eq(D.1)}. As a
result, the chiral hamiltonian (\ref{31}) has the fractional chiral
symmetries
\begin{equation}
\begin{tabular}{lll}
$\mathcal{C}_{x}H\left[ f_{x},g_{x};f_{y},g_{y};f_{z},g_{z}\right] \mathcal{C%
}_{x}^{-1}$ & $=$ & $H\left[ -f_{x},-g_{x};f_{y},g_{y};f_{z},g_{z}\right] $
\\
$\mathcal{C}_{y}H\left[ f_{x},g_{x};f_{y},g_{y};f_{z},g_{z}\right] \mathcal{C%
}_{y}^{-1}$ & $=$ & $H\left[ f_{x},g_{x};-f_{y},-g_{y};f_{z},g_{z}\right] $
\\
$\mathcal{C}_{z}H\left[ f_{x},g_{x};f_{y},g_{y};f_{z},g_{z}\right] \mathcal{C%
}_{z}^{-1}$ & $=$ & $H\left[ f_{x},g_{x};f_{y},g_{y};-f_{z},-g_{z}\right] $%
\end{tabular}
\label{te}
\end{equation}%
with chiral $\mathcal{C}$\ appearing as just their compositions. Moreover,
as for the chiral symmetry, these $\mathcal{C}_{i}$- symmetries leave
invariant the gap energy%
\begin{equation}
E_{g}=2\sqrt{\mathrm{f}_{x}^{2}+\mathrm{g}_{x}^{2}+\mathrm{f}_{y}^{2}+%
\mathrm{g}_{y}^{2}+\mathrm{f}_{z}^{2}+\mathrm{g}_{z}^{2}}  \label{ge}
\end{equation}%
For real $f_{i}$ and $g_{i}$, the vanishing of this gap energy requires $%
f_{x}=f_{y}=f_{z}=0$ and $g_{x}=g_{y}=g_{z}=0$. But these zeros have an
interpretation in terms of the chiral symmetry $\Gamma _{7}$ and of its
fractions $\mathcal{C}_{x},\mathcal{C}_{y}$ and $\mathcal{C}_{z}$; they are
just the fixed points of the chiral symmetry mapping $\left(
f_{i},g_{i}\right) $ into $\left( -f_{i},-g_{i}\right) $. At the \textrm{%
fixed} loci $\left( f_{i},g_{i}\right) =\left( 0,0\right) $, \textrm{one
obtains conditions on the k}$_{x},$\textrm{k}$_{y},$\textrm{k}$_{z}$\textrm{%
\ components of the momentum as well as on the parameters of the model. For
example, if taking }$f_{i}\sim \sin k_{i}$\textrm{, the solution }$f_{i}=0$%
\textrm{\ requires }$k_{i}=n_{i}\pi $ ($n_{i}=0$\textrm{\ or }$1);$ thus
\textrm{defining} stable points $\left( n_{x}\pi ,n_{y}\pi ,n_{z}\pi \right)
$ in the Brillouin Zone (BZ) as depicted by the Figure \textbf{\ref{d}}.
Like for \textrm{the} chiral symmetry $\Gamma _{7}=\mathcal{C}_{x}\mathcal{C}%
_{y}\mathcal{C}_{z}$, one may also talk about the fixed points of either the
2/3 symmetry $\mathcal{C}_{x}\mathcal{C}_{y}$ or the 1/3 $\mathcal{C}_{x}$
symmetries. In the 2/3 case, the fix locus of the $\mathcal{C}_{x}$ and $%
\mathcal{C}_{y}$ symmetries is%
\begin{equation}
\left( 0,0;0,0;f_{z},g_{z}\right)
\end{equation}%
on which the gap energy reduces to $E_{g}^{z}=2(\mathrm{f}_{z}^{2}+\mathrm{g}%
_{z}^{2})^{1/2}$. In this situation, one can formally talk about 1D state
propagating in the z-direction. For the 1/3 case, the locus of the $\mathcal{%
C}_{x}$ symmetry is
\begin{equation}
\left( 0,0;f_{y},g_{y};f_{z},g_{z}\right)
\end{equation}%
with gap energy as $E_{g}^{yz}=2(\mathrm{f}_{y}^{2}+\mathrm{g}_{y}^{2}+%
\mathrm{f}_{z}^{2}+\mathrm{g}_{z}^{2})^{1/2}$. Here, we have 2D states
propagating in the y-z plane. \textrm{\ }

\subsection{TRS and Mirror symmetries}

To describe the loci of the $\mathcal{C}_{i}$ fix points in BZ and the value
of the gap energy there, we have to know the local expressions of $f_{i}$
and $g_{i}$ as functions of momentum and coupling parameters. A manner to
fix these functions is to impose extra symmetries and use physical
arguments. In addition to the chiral symmetry $\mathcal{C}$, we demand two
more kinds of symmetries TRS and mirrors discussed below.

\subsubsection{Time reversal symmetry}

\begin{figure}[tbph]
\begin{center}
\hspace{0cm} \includegraphics[width=6cm]{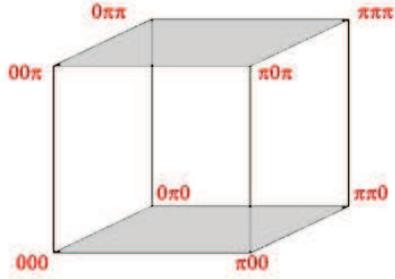}
\end{center}
\par
\vspace{-0.5cm}
\caption{The eight Dirac points in the reciprocal space given by the corners
of the cube. The grey faces are fix loci of the reflection $\mathcal{M}%
_{z}=i\Upsilon _{z}\Gamma _{7}.$}
\label{d}
\end{figure}

Time reversing symmetry TRS acts on hamiltonians like $\mathcal{T}H\left(
\mathbf{k};\mathbf{\Delta }\right) \mathcal{T}^{-1}=H\left( -\mathbf{k};%
\mathbf{\Delta }\right) $. As this symmetry combined with chiral $\mathcal{C}
$ is also a symmetry, we also have particle-hole symmetry $\mathcal{P}=$ $%
\mathcal{CT}$ acting as $\mathcal{P}H\left( \mathbf{k};\mathbf{\Delta }%
\right) \mathcal{P}^{-1}=-H\left( -\mathbf{k};\mathbf{\Delta }\right) $. If
moreover, we require that $\mathcal{T}^{2}=I_{id}$, we end up with the DBI
class of the AZ classification where $\mathcal{T}$ is realised by the
complex conjugation $K$ and then $\mathcal{P}$ as $\Gamma _{7}K$. So, a good
choice of the local functions $f_{i}=f\left( k_{i},\Delta _{i}\right) $ and $%
g_{i}=g\left( k_{i},\Delta _{i}\right) $ in the Brillouin Zone (BZ) can be
motivated by looking for 3D generalisation of known 1D topological systems
from AZ table having this specific symmetries like the 1D SSH theory \textrm{%
\cite{D6C,D6D}; }this leads to\textrm{\ }%
\begin{equation}
f_{i}=t_{i}\sin k_{i}\text{ \ },\qquad g_{i}=\Delta _{i}-t_{i}\cos k_{i}
\label{32}
\end{equation}%
where the $t_{i}$'s are the hopping parameters between the neighboring unit
cells and the $\Delta _{i}$'s are three coupling constants; they describe
the hoppings within a 3-dim unit cell. It is well noting that the eight band
Hamiltonian can be also expressed in the real space as reported in Appendix
C.\newline
Below, we set $t_{x}=t_{y}=t_{z}=1$ for simplicity of calculations; thus
reducing the couplings to the three $\Delta _{i}$'s; so we are left with a
total of 6 generic parameters namely the three $\left(
k_{x},k_{y},k_{z}\right) $ of momentum space and the three $\left( \Delta
_{x},\Delta _{y},\Delta _{z}\right) $ of the mass space. Notice the two
following features useful in the study of the spectrum of the hamiltonian (%
\ref{31}-\ref{32})\textrm{. }First, under time reversing symmetry $T$, we
have $f\left( -k_{i},\Delta _{i}\right) =-f\left( k_{i},\Delta _{i}\right) $%
\ while $g\left( -k_{i},\Delta _{i}\right) =g\left( k_{i},\Delta _{i}\right)
$; the fix points of this symmetry are given by the condition $f\left(
k_{i},\Delta _{i}\right) =0$ solved in the $\left( k_{x},k_{y},k_{z}\right) $
space by eight points $k_{i\ast }=0,$ $\pi $ mod 2$\pi $; these are the fix
points of $\mathcal{T}$ in the BZ; they are given by the eight corners of
the Figure \textbf{\ref{d}}.
Second, the real space image of Fig \textbf{\ref{d}} is given by the inverse
Fourier transforms $\tilde{f}\left( x_{i},\Delta _{i}\right) $ of the
functions $f\left( k_{i},\Delta _{i}\right) $; for the example of the x-
component, we have%
\begin{equation}
\tilde{f}_{x}=i\pi \left[ \delta \left( x-a\right) -\delta \left( x+a\right) %
\right]
\end{equation}%
with peaks at $x=\pm a$ mod 2a where $a\equiv 1$ is the lattice parameter in
x- direction. Similar relations can be also written down for the other
directions; and a picture like the Figure \textbf{\ref{d}} can be also drawn
in real space.

\subsubsection{Mirror symmetries}

With the choice (\ref{32}), the 3D hamiltonian thus obtained belongs to the
BDI family of the AZ classification \cite{F1,F2}; in addition to the CPT
symmetries, the Hamiltonian (\ref{31}) has also three reflections given by
\begin{equation}
\begin{tabular}{lll}
$\mathcal{M}_{x}H\left( k_{x},k_{y},k_{z};\mathbf{\Delta }\right) \mathcal{M}%
_{x}^{-1}$ & $=$ & $H\left( -k_{x},k_{y},k_{z};\mathbf{\Delta }\right) $ \\
$\mathcal{M}_{y}H\left( k_{x},k_{y},k_{z};\mathbf{\Delta }\right) \mathcal{M}%
_{y}^{-1}$ & $=$ & $H\left( k_{x},-k_{y},k_{z};\mathbf{\Delta }\right) $ \\
$\mathcal{M}_{z}H\left( k_{x},k_{y},k_{z};\mathbf{\Delta }\right) \mathcal{M}%
_{z}^{-1}$ & $=$ & $H\left( k_{x},k_{y},-k_{z};\mathbf{\Delta }\right) $%
\end{tabular}
\label{3m}
\end{equation}%
with the $\mathcal{M}_{i}$ mirrors realized as follows
\begin{equation}
\mathcal{M}_{x}=i\Upsilon _{x}\Gamma _{7}\text{\quad },\text{\quad }\mathcal{%
M}_{y}=i\Upsilon _{y}\Gamma _{7}\text{\quad },\text{\quad }\mathcal{M}%
_{z}=i\Upsilon _{z}\Gamma _{7}  \label{2m}
\end{equation}%
The fix points of these $\mathcal{M}_{i}$'s are given by faces of the Figure
\textbf{\ref{d}}; for example the fix points of $\mathcal{M}_{z}$ are given
by the surfaces $\left( k_{x},k_{y},0\right) $ and $\left( k_{x},k_{y},\pi
\right) $ represented by the grey surfaces of the Figure \textbf{\ref{d}}.
Eqs(\ref{3m}) teach us that the hamiltonian has other symmetries induced by
the composition of its basic symmetries namely $\mathcal{T}$, $\mathcal{C}%
_{i}$ and $\mathcal{M}_{i}$. For example, the composition of the $\mathcal{M}%
_{i}$'s amongst themselves like for inversion $\mathcal{I}$ reading as $i%
\mathcal{M}_{z}\mathcal{M}_{y}\mathcal{M}_{x}$ and realised by the generator
$i\Lambda _{2}\Lambda _{4}\Lambda _{6}$. The \textrm{common} denominator of
these operators and others is that they are symmetries of the gap energy (%
\ref{ge}). By substituting $k_{i\ast }=n_{i}\pi $ back into the expression
of the gap energy, we get%
\begin{equation}
E_{g}=2\sqrt{\left[ \Delta _{x}-\cos n_{x}\pi \right] ^{2}+\left[ \Delta
_{y}-\cos n_{y}\pi \right] ^{2}+\left[ \Delta _{z}-\cos n_{y}\pi \right] ^{2}%
}
\end{equation}%
which vanishes for $\Delta _{i}=\left( -\right) ^{n_{i}}$. At the Dirac
point $k_{x\ast }=k_{y\ast }=k_{z\ast }=0$, the lines $\Delta _{x}=\Delta
_{y}=\Delta _{z}=1$ in the moduli space $\mathcal{E}_{\Delta }$ define the
frontier between the non trivial topological phase $\left \vert \Delta
_{i}\right \vert <1$ and the trivial one given by $\left \vert \Delta
_{i}\right \vert >1$.

\section{Third order topological phase}

In this section, we study \textrm{the conditions for having higher order
topological phases in the chiral model described above and give their
solutions. }First, we give the continuum limits of the lattice model (\ref%
{31}) near the Dirac points and derive the HOT constraints. Then, we turn to
construct the gapless corner states.

\subsection{Continuum limit and Dirac hamiltonian}

Because of TRS symmetry, the 3D lattice model (\ref{32}) has eight Dirac
limits; the hamiltonians $H_{\mathbf{n}\pi }$ near these points are obtained
by expanding (\ref{32}) near $k_{i}=\kappa _{i\ast }+k_{i}^{\prime }$ (with $%
\kappa _{i\ast }=n_{i}\pi $), thus leading to the eight Dirac like
hamiltonians living on the corners of the Figure \textbf{\ref{d}}%
\begin{equation}
H_{\mathbf{n}\pi }=e^{in_{1}\pi }\Upsilon ^{x}\left( k_{x}^{\prime }-i\phi
_{x,n_{1}}\right) +e^{in_{2}\pi }\Upsilon ^{y}\left( k_{y}^{\prime }-i\phi
_{y,n_{2}}\right) +e^{in_{3}\pi }\Upsilon ^{z}\left( k_{z}^{\prime }-i\phi
_{z,n_{3}}\right)  \label{hn}
\end{equation}%
In this parametric expression, $\mathbf{n}$ is a vector integer $\left(
n_{1},n_{2},n_{3}\right) $ with $n_{i}=0,1$; and the $\phi _{l,n_{l}}$ are
matrices given by $g_{l,n_{l}}\mathcal{C}_{l}$ with $\mathcal{C}_{l}$ as in (%
\ref{tt}) and the $g_{l,n_{l}}$'s are eight mass like parameters given by $%
g_{l,n_{l}}=1-e^{in_{l}\pi }\Delta _{l}$ and living at the faces of the
Figure \textbf{\ref{d}}. For example $g_{x,0}=1-\Delta _{x}$ sits on the
face $\left( 0,k_{y},k_{z}\right) $ and $g_{x,\pi }=1+\Delta _{x}$ sits on $%
\left( \pi ,k_{y},k_{z}\right) $.

\subsubsection{Mass parameters and gap energy}

The eight Dirac hamiltonians (\ref{hn}) have the form $\Upsilon ^{x}K_{x,%
\mathbf{n}\pi }+\Upsilon ^{y}K_{y,\mathbf{n}\pi }+\Upsilon ^{z}K_{z,\mathbf{n%
}\pi }$; they have quite similar structure; they differ only by the value of
the $K_{l,\mathbf{n}\pi }$ coefficients in front of the gamma $\Upsilon _{l}$%
's; so their studies are quite similar and we can restrict the analysis to
one of them; say the Dirac limit near $\mathbf{k}_{\ast }=\left(
0,0,0\right) $ namely%
\begin{equation}
H_{\mathbf{0}}=\Upsilon ^{x}\left( k_{x}-i\phi _{x}\right) +\Upsilon
^{y}\left( k_{y}-i\phi _{y}\right) +\Upsilon ^{z}\left( k_{z}-i\phi
_{z}\right)  \label{h0}
\end{equation}%
with
\begin{equation}
\phi _{x}=g_{1,0}\mathcal{C}_{x}\quad ,\quad \phi _{y}=g_{2,0}\mathcal{C}%
_{y}\quad ,\quad \phi _{z}=g_{3,0}\mathcal{C}_{z}  \label{ff}
\end{equation}
where $g_{l,0}=1-\Delta _{l}$; and where we have dropped out the prime in $%
k_{i}^{\prime }$. Here, we want to give three comments regarding the two
above relations as they are at the basis in the study of HOT matter.

\  \

$\bullet $ \emph{Comment 1}: \emph{three} \emph{masses} $%
m_{_{x}},m_{_{y}},m_{_{z}}$\newline
If setting $k_{i}=0$ in (\ref{h0}), the eigenvalue equation $H_{\mathbf{0}%
}\Psi =E\Psi $ reduces to $-i\Upsilon ^{l}\phi _{l}\Psi =E\Psi $; and by
using (\ref{ff}), one can put it into the form $-i\Upsilon ^{l}g_{l,0}\left(
\mathcal{C}_{l}\Psi \right) =E\Psi $ with left hand side involving the
quantities $\mathcal{C}_{l}\Psi $. By thinking of the wave function as an
eigenstate of the fractional chiral charges, we can set $\mathcal{C}_{l}\Psi
=\theta _{l}\Psi $, and then bring the energy eigenvalue equation to the
form $-i\Upsilon ^{l}m_{l}\Psi =E\Psi $ with $m_{l}$ given by%
\begin{equation}
m_{x}=\theta _{_{x}}g_{x,0}\qquad ,\qquad m_{y}=\theta _{_{y}}g_{y,0}\qquad
,\qquad m_{z}=\theta _{z}g_{z,0}  \label{mx}
\end{equation}%
and the $\theta _{_{x}},\theta _{_{y}},\theta _{_{z}}$ charges as in eq(\ref%
{3q}). These relations indicate that good mass- like parameters are given by
$\left( m_{_{x}},m_{_{y}},m_{_{z}}\right) $ rather than $\left(
g_{x,0},g_{y,0},g_{z,0}\right) $. The dependence of the mass terms in $%
\theta _{_{i}}$'s is a premise manifestation of the fractionalisation of the
wave function.

\  \  \  \

\begin{figure}[tbph]
\begin{center}
\hspace{0cm} \includegraphics[width=8cm]{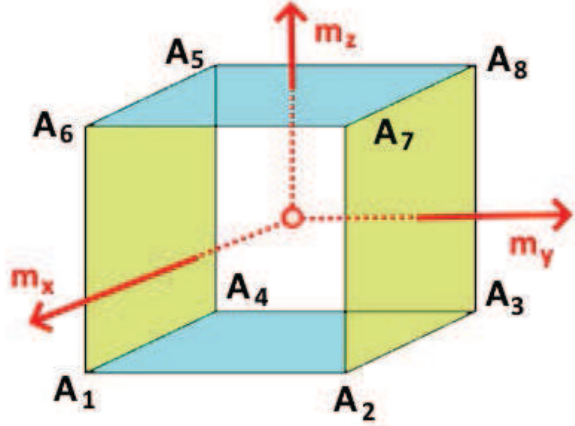}
\end{center}
\par
\vspace{-0.5cm}
\caption{The space $\mathcal{E}_{M}$ with variables $\left( \protect \mu _{x},%
\protect \mu _{y},\protect \mu _{z}\right) $ describing mass cube $\left \vert
\protect \mu _{i}\right \vert \leq g_{i,0}$. }
\label{m}
\end{figure}

$\bullet $ \emph{Comment 2}: \emph{mass cube and partial chiral symmetries}
\newline
The \textrm{three }$\theta _{i}$ charges of the partial chiral symmetry
operators $\mathcal{C}_{i}$ take the values $\pm 1$; by substituting these
values back into the triplet $\left( \theta _{_{x}},\theta _{_{y}},\theta
_{_{z}}\right) $, one generates 8 possibilities related by $\mathcal{C}_{i}$
transformations; the first one is $\left( +,+,+\right) ,$ the second reads
as $\left( -,-,-\right) $ and 6 more others ones; three of them have 2 $%
\left( +\right) $ and one $\left( -\right) $ like $\left( +,+,-\right) $,
and the three others have 2 $\left( -\right) $ and one $\left( +\right) $
like $\left( -,-,+\right) $. These 8 possibilities can be also splitted into
4+4 sets: 4 triplets with a product $\theta _{_{x}}\theta _{_{y}}\theta
_{_{z}}$ equals to +1; and 4 others with $\theta _{_{x}}\theta _{_{y}}\theta
_{_{z}}=-1$,
\begin{equation}
\begin{tabular}{lllllll}
$\left( m_{x}^{+},m_{y}^{+},m_{z}^{+}\right) $ & , & $\left(
m_{x}^{+},m_{y}^{-},m_{z}^{-}\right) $ & , & $\left(
m_{x}^{-},m_{y}^{+},m_{z}^{-}\right) $ & , & $\left(
m_{x}^{-},m_{y}^{-},m_{z}^{+}\right) $ \\
$\left( m_{x}^{-},m_{y}^{-},m_{z}^{-}\right) $ & , & $\left(
m_{x}^{-},m_{y}^{+},m_{z}^{+}\right) $ & , & $\left(
m_{x}^{+},m_{y}^{-},m_{z}^{+}\right) $ & , & $\left(
m_{x}^{+},m_{y}^{+},m_{z}^{-}\right) $%
\end{tabular}
\label{mm}
\end{equation}%
This classification is helpful when computing Ind(H), the topological index
of the hamiltonian. Notice that eq(\ref{mm}) is not sensitive to the signs
of $g_{i,0}$; so we assume below $g_{i,0}>0$ and denote the positive $m_{i}$%
's like $m_{i}^{+}=g_{i,0}$ and the negative ones as $m_{i}^{-}=-g_{i,0}$;
they vanish at the fix points of $\mathcal{C}_{i}$. Notice also that (\ref%
{mm}) are precisely the corners A$_{i}$ of the cube depicted by the Figure
\textbf{\ref{m}}; these points belong to the moduli space $\mathcal{E}_{M}$
with coordinates $\left( \mu _{x},\mu _{y},\mu _{z}\right) $ with $%
\left \vert \mu _{i}\right \vert \leq g_{i,0}$.
It is remarkable that despite the requirement $g_{i,0}>0,$ one do have
positive masses but also negative ones; they are needed by higher order
topology; in the limit $g_{z,0}\rightarrow 0$; the 3D cube given by the
Figure \textbf{\ref{m} }shrinks to the 2D olive-colored faces which merge;
by taking also the limit $g_{y,0}\rightarrow 0$, the corresponding face
shrinks to a 1D edge; and by performing moreover $g_{x,0}\rightarrow 0$, the
line shrinks to the central point of the cube where live the AZ gapless
state. From this view, the mass cube $\left \vert \mu _{i}\right \vert \leq
g_{i,0}$ is just an irregular blow up of the $\mathcal{C}$ fix point leading
to the fractionalisation of chiral symmetry as $\mathcal{C}_{x}\mathcal{C}%
_{y}\mathcal{C}_{z}$.

\  \  \

$\bullet $ \emph{Comment 3}: \emph{gap energy}\newline
At $k_{i}=n_{i}\pi $, \ the gap energy (\ref{eg}) can be expressed in two
different, but equivalent, forms; either like $%
2(g_{x,0}^{2}+g_{y,0}^{2}+g_{z,0}^{2})^{1/2}$ as usually done; or more
interestingly as follows%
\begin{equation}
E_{g}=2\sqrt{m_{x}^{2}+m_{y}^{2}+m_{z}^{2}}  \label{eg}
\end{equation}%
The two expressions are obviously equal since $m_{i}^{2}=\left( \theta
_{i}g_{i,0}\right) ^{2}$ is equal to $g_{i,0}^{2}$ due to the fact that $%
\theta _{i}=\pm 1$; the $E_{g}$ is not sensitive to the change of the sign
of the mass parameters; i.e to the replacement $m_{i}\rightarrow -m_{i}$; so
it is enough to study the situation at one corner of the Figure \textbf{\ref%
{m}}; say at $\left( m_{x}^{+},m_{y}^{+},m_{z}^{+}\right) $ and use the
above $\mathcal{C}_{i}$ symmetries to deduce the picture at the other
corners. Recall that for AZ matter, the topological transition takes place
at the fix point of $m_{i}\rightarrow -m_{i}$ corresponding to the vanishing
of masses; this is the fix point of the chiral symmetry $\Gamma _{7}$
thought of as $\mathcal{C}_{_{x}}\mathcal{C}_{_{y}}\mathcal{C}_{z}$.

\subsubsection{Higher order topological matter}

\textrm{We begin by recalling that 3D topological AZ matter has: }$\left(
1\right) $\textrm{\ two periodic boundaries and one open direction (say x-
and y- periodic but z- open). }$\left( 2\right) $\textrm{\ gapped states in
the 3D bulk but gapless surface states on the boundary. Regarding the 3D HOT
we are studying here, all the three real directions of the 3D material are
open and has gapless corners states; see Appendix \textbf{A}. To build the
gapless corner states}, we start from eqs(\ref{h0}-\ref{ff}), and proceed in
two steps as follows:\newline
First, we demand the bulk states of HOT matter to be gapped; that is $E_{g}>0
$; this is a necessary condition for higher order topological matter. A
direct implication of this constraint follows from eq(\ref{eg}); it
indicates that at least one of the three $m_{l}$ masses must be non
vanishing in order to have a non zero bulk gap. In terms of the $\Delta _{l}$
coupling constants, at least one of the three \ should be different from $1$%
. For a third order topological phase, the condition of non vanishing gap is
fulfilled by the non vanishing of the three masses; i.e $m_{l}\neq 0$; these
three conditions can be stated as follows
\begin{equation}
m_{x}m_{y}m_{z}\neq 0  \label{fx}
\end{equation}%
Second, we map (\ref{h0}) to the real space with coordinates $\mathbf{r}%
=\left( x,y,z\right) $ in order to exhibit explicitly the effect of
demanding the constraint (\ref{fx}); with this mapping the delta Dirac
distribution $\delta _{3}\left( \mathbf{k}-\mathbf{k}_{\ast }\right) $ turns
into a 3D space integral of the plane wave $e^{-i\left( \mathbf{k}-\mathbf{k}%
_{\ast }\right) .\mathbf{r}}$ and the vector $\mathbf{k}$ is represented by
the operator $\partial /i\partial \mathbf{r}$; then the hamiltonian involved
in the energy eigenvalue equation $H_{\mathbf{0}}\hat{\Psi}\left( \mathbf{r}%
\right) =E\hat{\Psi}\left( \mathbf{r}\right) $ reads as
\begin{equation}
H_{\mathbf{0}}=\Upsilon ^{x}\left( \frac{\partial }{i\partial x}-i\phi
_{x}\right) +\Upsilon ^{y}\left( \frac{\partial }{i\partial y}-i\phi
_{y}\right) +\Upsilon ^{z}\left( \frac{\partial }{i\partial z}-i\phi
_{z}\right)
\end{equation}%
Notice that besides its differential nature, $H_{\mathbf{0}}$ is still an 8$%
\times $8 matrix operator and $\hat{\Psi}\left( \mathbf{r}\right) $ is a
complex spinor with eight components. Notice also that the $\phi _{i}$'s are
matrix masses that have to be treated carefully when looking for solution of
the eigenstates of $H_{\mathbf{0}}$. To engineer gapless states while
preserving the constraint (\ref{fx}), we look for exotic (non oscillating)
wave functions $\Psi \left( \mathbf{r}\right) $ solving the equation $H_{%
\mathbf{0}}\Psi \left( \mathbf{r}\right) =0$. Clearly a solution of such
zero mode equation exists and has the form of an evanescent wave
\begin{equation}
\Psi \left( \mathbf{r}\right) =\mathcal{N}e^{-\mathbf{m}.\left( \mathbf{r-r}%
_{0}\right) }\Omega   \label{wf}
\end{equation}%
with vector $\mathbf{m}$ referring to $\left( m_{x},m_{y},m_{z}\right) $ and
where $\mathbf{r}_{0}$ is some given space position; \textrm{it is imagined
as a corner}. The use of $\mathbf{m}$ instead of the $\phi _{i}$'s is
because of their relationship (\ref{mx}) and also because of (\ref{mm}).
\begin{figure}[tbph]
\begin{center}
\hspace{0cm} \includegraphics[width=10cm]{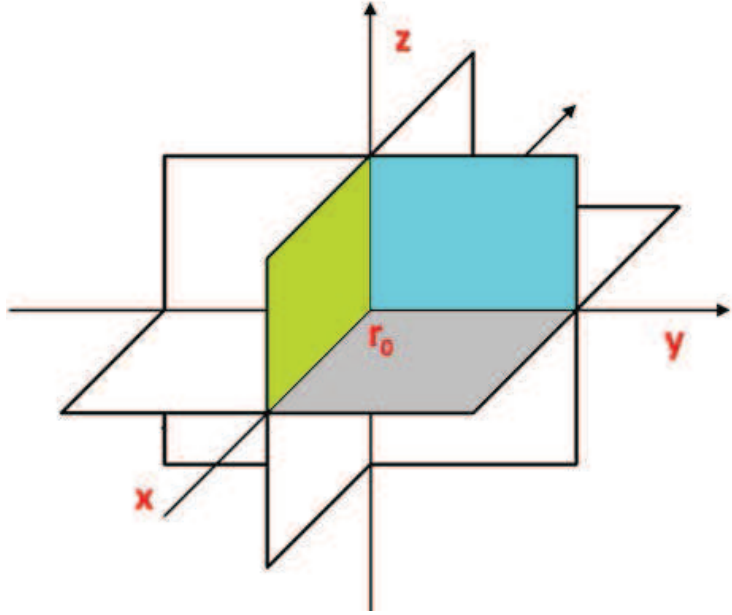}
\end{center}
\par
\vspace{-0.5cm}
\caption{An octant of the 3D space $\mathbb{R}^{3}$ with corner at $\mathbf{r%
}_{0}$; it is given by $\frac{1}{8}$ of $\mathbb{R}^{3}$. Octants of $%
\mathbb{R}^{3}$ are defined by the signs of $x-x_{0},$ $y-y_{0}$ and $%
z-z_{0}.$}
\label{oc}
\end{figure}
The factor $\mathcal{N}$ is a normalisation constant and the factor $\Omega $
is a constant spinor with eight component; it gives the polarisation of $%
\Psi \left( \mathbf{r}\right) $. Notice also that the existence of the
candidate solution (\ref{wf}) can be also motivated from the gap energy
formula $E_{g}=2\sqrt{m_{j}^{2}+k_{j}^{2}}$ which never vanishes as far as (%
\ref{fx}) holds and as long as the $k_{j}$'s are real numbers. However, a
vanishing $E_{g}$ is possible for pure imaginary momenta $k_{j}=im_{j}$
indicating that gapless states are given by corner waves.

\subsection{Gapless corner states}

Here, we sit in a particular region in the moduli space $\mathcal{E}_{M}$
with coordinates $\left( \mu _{x},\mu _{y},\mu _{z}\right) $; in addition to
the constraint (\ref{fx}), we assume moreover
\begin{equation}
m_{x}>0\quad ,\quad m_{y}>0\quad ,\quad m_{z}>0  \label{as}
\end{equation}%
This choice corresponds to sitting at the $\left(
m_{x}^{+},m_{y}^{+},m_{z}^{+}\right) $ point of the set (\ref{mm}); it is
one of eight possibilities corresponding to the eight corners of the Figure
\textbf{\ref{m}}; the above (\ref{as}) corresponds to the corner \textrm{A7}%
. To that purpose, we first derive the solution (\ref{wf}) describing
gapless corner states; then we turn to give some of their characteristic
properties.

\subsubsection{Deriving Eq(\protect \ref{wf})}

We begin by expressing the gapless state equation $H_{\mathbf{k}}\Psi =0$ as%
\begin{equation}
\left( -i\mathbf{\Upsilon }^{x}\mathcal{D}_{x}-i\mathbf{\Upsilon }^{y}%
\mathcal{D}_{y}-i\mathbf{\Upsilon }^{z}\mathcal{D}_{z}\right) \Psi =0
\label{kh}
\end{equation}%
with%
\begin{equation}
\mathcal{D}_{i}=\partial _{i}-\phi _{i}  \label{3d}
\end{equation}%
and $\phi _{i}$ mass matrices as%
\begin{equation}
\phi _{x}=g_{x}\mathcal{C}_{x}\quad ,\quad \phi _{y}=g_{y}\mathcal{C}%
_{y}\quad ,\quad \phi _{z}=g_{z}\mathcal{C}_{z}  \label{d3}
\end{equation}%
where the $\mathcal{C}_{i}$'s are given by (\ref{tt}). Since these $\mathcal{%
C}_{i}$'s are hermitian matrices and they commute; i.e: $\mathcal{C}_{i}%
\mathcal{C}_{j}=\mathcal{C}_{j}\mathcal{C}_{i}$; it results that the $\phi
_{i}$'s are hermitian as well and they commute between them. So, the
solution of (\ref{kh}) can be factorised as%
\begin{equation}
\Psi \left( x,y,z\right) =\mathcal{N}e^{-\phi _{x}\left( x-x_{0}\right)
}e^{-\phi _{y}\left( y-y_{0}\right) }e^{-\phi _{z}\left( z-z_{0}\right)
}\Omega  \label{om}
\end{equation}%
with $\Omega $ a constant unit spinor with components $\left( \omega
_{1},...,\omega _{8}\right) $. By constant spinor $\Omega $, we mean that it
is space coordinate independent like $\partial \Omega /\partial x=0$ and so
on; and by unit spinor we refer to the normalisation $\left \vert \omega
_{1}\right \vert ^{2}+...+\left \vert \omega _{8}\right \vert ^{2}=1.$ These
two specific features can be stated as follows:%
\begin{equation}
\partial _{i}\Omega =0\quad ,\quad tr\left( \Omega ^{+}\Omega \right) =1
\end{equation}%
Using these properties, we can compute the normalisation factor $\mathcal{N}$
by using the relation $\int_{\mathbb{R}^{3}}tr\left( \Psi ^{+}\Psi \right)
d^{3}\mathbf{r}=1$. But to perform this calculation, we need to know the
value of $tr\left( \Psi ^{+}\Psi \right) $ which requires the introduction
of the charge vector $\boldsymbol{q}$ with components $%
(q_{_{x}},q_{_{y}},q_{_{z}})$; this need of $\boldsymbol{q}$ is because of
the matrix masses $\phi _{i}=g_{i}\mathcal{C}_{i}$ appearing in the
exponentials of eq(\ref{om}). So, normalisable solutions depend on the
values of the charge vector $\boldsymbol{q}$ as shown below%
\begin{equation}
\Psi _{\boldsymbol{q}}=\mathcal{N}e^{-m_{x}\left( x-x_{0}\right)
}e^{-m_{y}\left( y-y_{0}\right) }e^{-m_{z}\left( z-z_{0}\right) }\Omega _{%
\boldsymbol{q}}
\end{equation}%
where we have used (\ref{mx}) and the eigenstate equation $\mathcal{C}%
_{i}\Omega _{\boldsymbol{q}}=\theta _{i}\Omega _{\boldsymbol{q}}$ whose
solution were studied in \textrm{subsection 2.2}. Depending on the sign of $%
(m_{_{x}},m_{_{y}},m_{_{z}})$, the normalisation of $\Psi _{\boldsymbol{q}}$
requires $x-x_{0}$, $y-y_{0}$ and $z-z_{0}$ having respectively the same
sign as the sign of $m_{_{x}},m_{_{y}},m_{_{z}}$. In this case, we have
\begin{equation}
\mathcal{N}=\frac{1}{\sqrt{8\left \vert m_{x}m_{y}m_{z}\right \vert }}
\end{equation}%
Notice that in the case where $sgn\left( m_{i}\right) $ are positive, we
must have
\begin{equation}
x>x_{0}\qquad ,\qquad y>y_{0}\qquad ,\qquad z>z_{0}
\end{equation}%
These three inequalities define an octant of the 3D space $\mathbb{R}^{3}$,
the three line sides intersect at the point $\left( x_{0},y_{0},z_{0}\right)
$ giving the corner of the octant. The relationship between the signs of $%
(m_{x},m_{y},m_{z})$ and the corners of the Figure \textbf{\ref{m}} is given
by the following table%
\begin{equation*}
\begin{tabular}{|l|l|l|l|l|l|l|l|l|}
\hline
${\small m_{x},m_{y},m_{z}}$ & ${\small +--}$ & ${\small ++-}$ & ${\small -+-%
}$ & ${\small ---}$ & ${\small --+}$ & ${\small +-+}$ & ${\small +++}$ & $%
{\small -++}$ \\ \hline
{\small Fig} \textbf{\ref{m}} & \ {\small A1} & {\small A2} & {\small A3} &
{\small A4} & {\small A5} & {\small A6} & {\small A7} & {\small A8} \\ \hline
\end{tabular}%
\end{equation*}

\subsubsection{Chiral wave functions}

A gapless chiral wave function $\Psi $ obeys the zero mode equation $H\Psi
=0 $ and moreover carries a definite charge $q_{_{\Gamma _{7}}}$ under the
chiral operator $\Gamma _{7}.$ Since $q_{_{\Gamma _{7}}}$ can take the
values $\pm 1,$ we distinguish two\ types of chiral wave spinors: right
handed $\Psi _{R}$ with $q_{_{\Gamma _{7}}}=+1$ and a left handed $\Psi _{L}$
with $q_{_{\Gamma _{7}}}=-1$. In other words, gapless \textrm{right and left}
handed wave functions are defined as%
\begin{equation}
\begin{tabular}{lll}
$H\Psi _{R}=0$ & $\qquad ,\qquad $ & $\Gamma _{7}\Psi _{R}=+\Psi _{R}$ \\
$H\Psi _{L}=0$ & $\qquad ,\qquad $ & $\Gamma _{7}\Psi _{L}=-\Psi _{L}$%
\end{tabular}%
\end{equation}%
The solutions of the chirality conditions are given by eqs(\ref{pr}) and (%
\ref{pl}) showing that they are combinations of four terms. Below, we show
that the eight possible polarisations of the $\Psi _{R}/\Psi _{L}$ states
will be dispatched on the corners of the mass cube according to the
values{}of the charges $\theta _{i}$ of the observables $\mathcal{C}_{i}$.
For that, notice that the polarisation of the gapless wave function $\Psi $ (%
\ref{om}) is carried by the constant spinor $\Omega $; and because of (\ref%
{d3}), it results that gapless right handed and left handed states are given
by
\begin{equation}
\begin{tabular}{lll}
$\Psi _{\mathbf{q,}R}$ & $=$ & $\frac{1}{\sqrt{8\left \vert
m_{x}m_{y}m_{z}\right \vert }}e^{-m_{x}\left( x-x_{0}\right)
}e^{-m_{y}\left( y-y_{0}\right) }e^{-m_{z}\left( z-z_{0}\right) }\Omega _{%
\mathbf{q,}R}$ \\
$\Psi _{\mathbf{q,}L}$ & $=$ & $\frac{1}{\sqrt{8\left \vert
m_{x}m_{y}m_{z}\right \vert }}e^{-m_{x}\left( x-x_{0}\right)
}e^{-m_{y}\left( y-y_{0}\right) }e^{-m_{z}\left( z-z_{0}\right) }\Omega _{%
\mathbf{q,}L}$%
\end{tabular}
\label{rl}
\end{equation}%
with vector $\mathbf{q}$ referring to $(q_{_{x}},q_{_{y}},q_{_{z}})$ and
polarisations $\Omega _{\mathbf{q,}R}$ and $\Omega _{\mathbf{q,}L}$ as
follows%
\begin{equation}
\Gamma _{7}\Omega _{\mathbf{q,}R}=\Omega _{\mathbf{q,}R}\qquad ,\qquad
\Gamma _{7}\Omega _{\mathbf{q,}L}=-\Omega _{\mathbf{q,}L}  \label{ch}
\end{equation}%
The chiral wave functions $\Psi _{R}$ and $\Psi _{L}$ have been
fractionalized into $\Psi _{\mathbf{q,}R}$ and $\Psi _{\mathbf{q,}L}$.
Notice also that in the limit $m_{i}\rightarrow 0,$ the fractional waves $%
\Psi _{\mathbf{q,}R/R}$ tend to $\Omega _{\mathbf{q,}R}\mathbf{\delta }%
_{3}\left( \mathbf{r}-\mathbf{r}_{0}\right) $ where $\mathbf{\delta }_{3}$
is the usual 3D Dirac delta function. Below, we determine these gapless
chiral waves and give their properties.

\section{Right and left handed states}

In this section, we give specific properties of fractional chiral states.
First, we focus on the study of gapless states with fractional right and
left handed chiralities. Then, we use these fractional chiral states to
calculate the topological index of the eight band model.

\subsection{Chiral polarisations $\Omega _{\mathbf{q,}R}$ and $\Omega _{%
\mathbf{q,}L}$}

From eqs(\ref{rl}), we learn that the fractional chiral wave functions $\Psi
_{\mathbf{q,}R}$ and $\Psi _{\mathbf{q,}L}$ are given by the product of two
factors: the extrinsic factor $\mathcal{N}e^{-\mathbf{m}\left( \mathbf{r}-%
\mathbf{r}_{0}\right) }$ shared by both waves; and intrinsic factors given
by $\Omega _{\mathbf{q,}R}$ and $\Omega _{\mathbf{q,}L}$ carrying their
polarisations. The explicit expressions of these polarisations are derived
below.

\subsubsection{Right handed polarisation}

To study the properties of the gapless right handed chirality of $\Omega _{%
\mathbf{q,}R}$, we have to look for the allowed possibilities of the
constant spinors $\Omega _{\mathbf{q,}R}$ solving (\ref{ch}). For that, we
use the factorized representation (\ref{g7}) to express the wave function
polarisation $\Omega _{\mathbf{q}}$ as follows%
\begin{equation}
\Omega _{\mathbf{q}}=\mathbf{a}_{q_{_{\varrho _{{\small 3}}}}}\mathbf{b}%
_{q_{_{\tau _{{\small 3}}}}}\mathbf{c}_{q_{_{\sigma _{{\small 3}}}}}
\label{ab}
\end{equation}%
where $\mathbf{a}_{\pm },\mathbf{b}_{\pm }$ and $\mathbf{c}_{\pm }$ are two-
component constant spinors.normalised as%
\begin{equation}
\mathbf{e}_{+}=\left(
\begin{array}{c}
1 \\
0%
\end{array}%
\right) \qquad ,\qquad \mathbf{e}_{-}=\left(
\begin{array}{c}
0 \\
1%
\end{array}%
\right)
\end{equation}%
with $\mathbf{e}_{\pm }$\textbf{\ }standing for $\mathbf{a}_{\pm },\mathbf{b}%
_{\pm }$ and $\mathbf{c}_{\pm }$. Recall that the \textbf{$\theta $} vector
charge $(\theta _{_{x}},\theta _{_{y}},\theta _{_{z}})$ is related to the $%
\mathbf{q}$ charge vector as $(-q_{_{\sigma _{{\small 3}}}},q_{_{\tau _{%
{\small 3}}}}q_{_{\sigma _{{\small 3}}}},q_{_{\varrho _{{\small 3}}}})$. By
help of (\ref{g7}), the chirality of $\Omega _{\mathbf{q}\text{,}R}$ can be
deduced from the value of $q_{_{\mathbf{\Gamma }_{7}}}=-q_{_{\varrho
_{3}}}q_{_{\tau _{3}}}$ which indicates that right handed wave function must
have $q_{_{\varrho _{3}}}q_{_{\tau _{3}}}=-1$; but $q_{_{\sigma _{{\small 3}%
}}}$ free that is taking both values $\pm 1$. As the condition $q_{_{\varrho
_{3}}}q_{_{\tau _{3}}}=-1$ has two solutions namely $\left( q_{_{\varrho
_{3}}},q_{_{\tau _{3}}}\right) $ equals either to $\left( 1,-1\right) $ or
to $\left( -1,1\right) $; it follows that there are four solutions
classified by the Pauli charge vectors
\begin{equation}
\mathbf{q}=\left( 1,-1,\pm 1\right) \qquad ,\qquad \left( -1,1,\pm 1\right)
\end{equation}%
Therefore, we have the four following $\Omega _{\mathbf{q},R}$
configurations
\begin{equation}
\begin{tabular}{|l|l|l|l|}
\hline
$\left( q_{_{\varrho _{3}}},q_{_{\tau _{3}}},q_{_{\sigma _{{\small 3}%
}}}\right) $ & $\Omega _{\mathbf{q},R}$ & $q_{_{\mathbf{\Gamma }_{7}}}$ & $%
(\theta _{_{x}},\theta _{_{y}},\theta _{_{z}})$ \\ \hline
$\left( -1,+1,+1\right) $ & $\mathbf{a}_{-}\mathbf{b}_{+}\mathbf{c}_{+}$ & $%
+1$ & $\left( -1,+1,-1\right) $ \\ \hline
$\left( +1,-1,+1\right) $ & $\mathbf{a}_{+}\mathbf{b}_{-}\mathbf{c}_{+}$ & $%
+1$ & $\left( -1,-1,+1\right) $ \\ \hline
$\left( -1,+1,-1\right) $ & $\mathbf{a}_{-}\mathbf{b}_{+}\mathbf{c}_{-}$ & $%
+1$ & $\left( +1,-1,-1\right) $ \\ \hline
$\left( +1,-1,-1\right) $ & $\mathbf{a}_{+}\mathbf{b}_{-}\mathbf{c}_{-}$ & $%
+1$ & $\left( +1,+1,+1\right) $ \\ \hline
\end{tabular}
\label{327}
\end{equation}%
\begin{equation*}
\end{equation*}%
However, not all these $\Omega _{\mathbf{q},R}$ configurations can be
substituted into (\ref{rl}); this is because of the correlation between that
the mass vector $\mathbf{m}$ and the position vector $\left( \mathbf{r-r}%
_{0}\right) $ as the convergence of tr$\left( \Psi _{\mathbf{q,}R}^{+}\Psi _{%
\mathbf{q,}R}\right) $ requires
\begin{equation}
\theta _{x}\left( x-x_{0}\right) >0\qquad ,\qquad \theta _{y}\left(
y-y_{0}\right) >0\qquad ,\qquad \theta _{z}\left( z-z_{0}\right) >0
\end{equation}%
Recall that tr$\left( \Psi _{\mathbf{q,}R}^{+}\Psi _{\mathbf{q,}R}\right) $
is equal to $\mathcal{N}e^{-2m_{x}\left( x-x_{0}\right) }e^{-2m_{y}\left(
y-y_{0}\right) }e^{-2m_{z}\left( z-z_{0}\right) }$ times tr$\left( \Omega _{%
\mathbf{q,}R}^{+}\Omega _{\mathbf{q,}R}\right) =1$. In these exponentials,
the masses are given by $m_{i}=\theta _{i}g_{i,0}$ as one sees on eq(\ref{as}%
), and because of our assumption $g_{i,0}>0,$ it results that convergence of
tr$\left( \Psi _{\mathbf{q,}R}^{+}\Psi _{\mathbf{q,}R}\right) $ requires $%
\theta _{x}$ and $\left( x-x_{0}\right) $ to have the same sign so that $%
e^{-2m_{x}\left( x-x_{0}\right) }$ vanishes when $m_{x}\left( x-x_{0}\right)
\rightarrow +\infty $. For the same reason, it requires also that $\theta
_{y}$ and $\left( y-y_{0}\right) $ should have the same sign and similarly
for $\theta _{z}$ and $\left( z-z_{0}\right) $. As such, according to the
signs of $(\theta _{_{x}},\theta _{_{y}},\theta _{_{z}}),$ normalised
gapless states $\Psi _{\mathbf{q,}R}$ are given by%
\begin{equation}
\begin{tabular}{|l|l|l|l|l|l|}
\hline
{\small Fig} \textbf{\ref{m}} & $sgn({\small m}_{_{x}}{\small ,m}_{_{y}}%
{\small ,m}_{_{z}})$ & ${\small sgn\tilde{x}}$ & ${\small sgn\tilde{y}}$ & $%
{\small sgn\tilde{z}}$ & $\Omega _{\mathbf{q},R}$ \\ \hline
{\small A1} & {\small \  \ }$\  \ (+,-,-)$ & {\small \ }$\  \ +$ & {\small \ }$%
\  \ -$ & {\small \ }$\  \ -$ & $\mathbf{a}_{-}\mathbf{b}_{+}\mathbf{c}_{-}$
\\ \hline
{\small A3} & {\small \  \ }$\  \ (-,+,-)$ & {\small \ }$\  \ -$ & {\small \ }$%
\  \ +$ & {\small \ }$\  \ -$ & $\mathbf{a}_{-}\mathbf{b}_{+}\mathbf{c}_{+}$
\\ \hline
{\small A5} & {\small \  \ }$\  \ (-,-,+)$ & {\small \ }$\  \ -$ & {\small \ }$%
\  \ -$ & {\small \ }$\  \ +$ & $\mathbf{a}_{+}\mathbf{b}_{-}\mathbf{c}_{+}$
\\ \hline
{\small A7} & {\small \  \ }$\  \ (+,+,+)$ & {\small \ }$\  \ +$ & {\small \ }$%
\  \ +$ & {\small \ }$\  \ +$ & $\mathbf{a}_{+}\mathbf{b}_{-}\mathbf{c}_{-}$
\\ \hline
\end{tabular}
\label{328}
\end{equation}%
where ${\small \tilde{x}}=x-x_{0}$\ and so on. Gapless right handed states
occupy four of the eight corners of the Figure \textbf{\ref{m}}; so one
expects that the other four corners are occupied by gapless left handed
states; this feature is shown below.

\subsubsection{Left handed polarisation}

Gapless left handed wave function $\Psi _{L}$ can be obtained by repeating
the above steps of calculations. The constraint $\Gamma _{7}\Omega
_{L}=-\Omega _{L}$ requires $q_{_{\mathbf{\Gamma }_{7}}}=-1$ and leads to
the four following $\Omega _{\mathbf{q},L}$ configurations
\begin{equation}
\begin{tabular}{|l|l|l|l|}
\hline
${\small (q}_{_{\varrho _{{\small 3}}}}{\small ,q}_{_{\tau _{{\small 3}}}}%
{\small ,q_{_{\sigma _{3}}})}$ & $\Omega _{\mathbf{q},L}$ & $q_{_{\mathbf{%
\Gamma }_{7}}}$ & $(\theta _{_{x}},\theta _{_{y}},\theta _{_{z}})$ \\ \hline
$\left( {\small +1,+1,+1}\right) $ & $\mathbf{a}_{+}\mathbf{b}_{+}\mathbf{c}%
_{+}$ & $-1$ & $\left( {\small -1,+1,+1}\right) $ \\ \hline
$\left( {\small -1,-1,+1}\right) $ & $\mathbf{a}_{-}\mathbf{b}_{-}\mathbf{c}%
_{+}$ & $-1$ & $\left( {\small -1,-1,-1}\right) $ \\ \hline
$\left( {\small +1,+1,-1}\right) $ & $\mathbf{a}_{+}\mathbf{b}_{+}\mathbf{c}%
_{-}$ & $-1$ & $\left( {\small +1,-1,+1}\right) $ \\ \hline
$\left( {\small -1,-1,-1}\right) $ & $\mathbf{a}_{-}\mathbf{b}_{-}\mathbf{c}%
_{-}$ & $-1$ & $\left( {\small +1,+1,-1}\right) $ \\ \hline
\end{tabular}%
\end{equation}%
The locations of these gapless left handed states in the mass space $%
\mathcal{E}_{M}$ are collected in the following table%
\begin{equation}
\begin{tabular}{|l|l|l|l|l|l|}
\hline
{\small Fig} \textbf{\ref{m}} & {\small sgn}${\small (m}_{{\small x}}{\small %
,m}_{_{y}}{\small ,m}_{_{z}}{\small )}$ & ${\small sgn\tilde{x}}$ & ${\small %
sgn\tilde{y}}$ & ${\small sgn\tilde{z}}$ & $\Omega _{\mathbf{q},L}$ \\ \hline
{\small A2} & {\small \  \ }$\  \ (+,+,-)$ & {\small \ }$+$ & {\small \ }$+$ &
$\ -$ & $\mathbf{a}_{-}\mathbf{b}_{-}\mathbf{c}_{-}$ \\ \hline
{\small A4} & {\small \  \ }$\  \ (-,-,-)$ & {\small \ }$-$ & $\ -$ & $\ -$ & $%
\mathbf{a}_{-}\mathbf{b}_{-}\mathbf{c}_{+}$ \\ \hline
{\small A6} & {\small \  \ }$\  \ (+,-,+)$ & {\small \ }$+$ & $\ -$ & $\ +$ & $%
\mathbf{a}_{+}\mathbf{b}_{+}\mathbf{c}_{-}$ \\ \hline
{\small A8} & {\small \  \ }$\  \ (-,+,+)$ & {\small \ }$-$ & {\small \ }$+$ &
{\small \ }$+$ & $\mathbf{a}_{+}\mathbf{b}_{+}\mathbf{c}_{+}$ \\ \hline
\end{tabular}
\label{330}
\end{equation}%
\begin{equation*}
\end{equation*}%
they occupy precisely the remaining corners in table (\ref{328}).

\subsection{Computing the topological index}

In this subsection, we calculate the topological index (IndH) of the chiral
hamiltonian H$_{\mathbf{k}}$ of the eight bands lattice model. This is an
integer number given by the difference $n_{R}-n_{L}$ counting the number of
chiral gapless states of the higher order topological model. To determine
IndH, we apply the index theorem on open spaces \cite{G1,G2,G3,G4,G5} to the
hamiltonian (\ref{31}) and use a practical formula obtained recently in
\textrm{\cite{nous}}. To that purpose, we start from (\ref{31}) with
coefficients $f_{i}=f\left( k_{i}\right) $ and $g_{i}=g\left( k_{i},\Delta
_{i}\right) $ as in (\ref{32}) and gap energy relation as in (\ref{eg}); the
continuum limit of (\ref{31}) leads to eight Dirac- like hamiltonians H$_{%
\mathbf{n\pi }}$ sitting at the eight corners of BZ represented by the
Figure \textbf{\ref{d}}. At each one of these corners, the gap energy $E_{g}$
has the same expression which is given by $2(k_{i}^{2}+m_{i}^{2})^{1/2}$;
thanks to mirror $\mathcal{M}_{l}$ and the fractional chiral symmetries $%
\mathcal{C}_{l}$. So, gapless states are of two kinds: First massless states
with $k_{i}=0$, but this is not our case. Second, massive states sitting at
the eight corners of the cube of the Figure \textbf{\ref{m}}; these gapless
states are described by evanescent waves with fractional chiral
polarisation; they are given by wave function type $\left \vert
m_{x}m_{y}m_{z}\right \vert ^{-1/2}e^{-\mathbf{m.\tilde{r}}}\Omega _{\mathbf{%
q}}$ with definite polarisation $\Omega _{\mathbf{q}}$ characterising the
HOT matter. By combining the two Figures \textbf{\ref{d}} and \textbf{\ref{m}%
}, one gets a total number of 64 massive corner states in the chiral lattice
model (\ref{31}). Consequently, IndH is equal to 64 times the contribution
of a corner state to the topological index which turns out to be given by
the unit $\dprod \nolimits_{i}\frac{1}{2}\left[ sgn(m_{i}^{+})-sgn(m_{i}^{-})%
\right] $. This result can be also derived by first expressing IndH as the
sum $\sum_{\mathbf{n}}$Ind(H$_{\mathbf{n\pi }}$) with integer vector $%
\mathbf{n}=\left( n_{x},n_{y},n_{z}\right) $ and $n_{i}=0,1$. Then, using
the property that the eight Dirac- like hamiltonians $H_{\mathbf{n\pi }}$
have same ground state properties as $H_{\mathbf{0\pi }}$; this leads to
IndH equals to 8 times Ind(H$_{\mathbf{0\pi }}$). To calculate the explicit
value of Ind(H$_{\mathbf{0\pi }}$), we use the formula of \textrm{\cite{nous}
according to which the topological index for cube like geometries is given
by }sgn functions as follows
\begin{equation}
\begin{tabular}{lll}
Ind(H$_{\mathbf{0\pi }}$) & $=$ & $+\frac{1}{8}\left[ sgn\left(
a_{+}b_{+}c_{+}\right) +sgn\left( a_{-}b_{-}c_{+}\right) \right] $ \\
&  & $+\frac{1}{8}\left[ sgn\left( a_{+}b_{-}c_{-}\right) +sgn\left(
a_{-}b_{+}c_{-}\right) \right] $ \\
&  & $-\frac{1}{8}\left[ sgn\left( a_{+}b_{-}c_{+}\right) +sgn\left(
a_{-}b_{+}c_{+}\right) \right] $ \\
&  & $-\frac{1}{8}\left[ sgn\left( a_{+}b_{+}c_{-}\right) +sgn\left(
a_{-}b_{-}c_{-}\right) \right] $%
\end{tabular}
\label{r}
\end{equation}%
To apply this formula to our situation, we think of the eight triplets $%
\left( a_{\pm },b_{\pm },c_{\pm }\right) $ appearing in above relations as
just the mass triplets $\left( m_{x}^{\pm },m_{y}^{\pm },m_{z}^{\pm }\right)
$ defining the coordinates of the eight corner points A$_{1}$,...,A$_{8}$ of
the Figure \textbf{\ref{m}}. For the case of corners A1- A3- A5- A7, the
associated sgn functions are learnt from the table (\ref{328}), thus leading
to%
\begin{equation}
\begin{tabular}{|l|l|l|l|l|}
\hline
{\small Fig} \textbf{\ref{m}} & $({\small m}_{_{x}}{\small ,m}_{_{y}}{\small %
,m}_{_{z}})$ & $sgn({\small m_{_{x}}m}_{{\small y}},{\small m}_{_{z}})$ & $%
\Omega _{\mathbf{q},R}$ & {\small Ind(H}$_{\mathbf{0\pi }}${\small )} \\
\hline
{\small A1} & {\small \  \ }$\  \ (+,-,-)$ & {\small \  \ }$\  \ +$ & $\mathbf{a}%
_{-}\mathbf{b}_{+}\mathbf{c}_{-}$ & $\frac{1}{8}$ \\ \hline
{\small A3} & {\small \  \ }$\  \ (-,+,-)$ & {\small \  \ }$\  \ +$ & $\mathbf{a}%
_{-}\mathbf{b}_{+}\mathbf{c}_{+}$ & $\frac{1}{8}$ \\ \hline
{\small A5} & {\small \  \ }$\  \ (-,-,+)$ & {\small \  \ }$\  \ +$ & $\mathbf{a}%
_{+}\mathbf{b}_{-}\mathbf{c}_{+}$ & $\frac{1}{8}$ \\ \hline
{\small A7} & {\small \  \ }$\  \ (+,+,+)$ & {\small \  \ }$\  \ +$ & $\mathbf{a}%
_{+}\mathbf{b}_{-}\mathbf{c}_{-}$ & $\frac{1}{8}$ \\ \hline
\end{tabular}%
\end{equation}%
Similarly, the sgn functions associated A2- A4- A6- A8 are read from the
table (\ref{330}) leading to,
\begin{equation}
\begin{tabular}{|l|l|l|l|l|}
\hline
{\small Fig} \textbf{\ref{m}} & ${\small sgn(m}_{{\small x}}{\small ,m}%
_{_{y}}{\small ,m}_{_{z}}{\small )}$ & ${\small sgn(m}_{_{x}}{\small m}_{%
{\small y}}{\small ,m}_{_{z}}{\small )}$ & $\Omega _{\mathbf{q},L}$ &
{\small Ind(H}$_{\mathbf{0\pi }}${\small )} \\ \hline
{\small A2} & {\small \  \ }$\  \ (+,+,-)$ & {\small \  \ }$\  \ -$ & $\mathbf{a}%
_{-}\mathbf{b}_{-}\mathbf{c}_{-}$ & $\frac{1}{8}$ \\ \hline
{\small A4} & {\small \  \ }$\  \ (-,-,-)$ & {\small \  \ }$\  \ -$ & $\mathbf{a}%
_{-}\mathbf{b}_{-}\mathbf{c}_{+}$ & $\frac{1}{8}$ \\ \hline
{\small A6} & {\small \  \ }$\  \ (+,-,+)$ & {\small \  \ }$\  \ -$ & $\mathbf{a}%
_{+}\mathbf{b}_{+}\mathbf{c}_{-}$ & $\frac{1}{8}$ \\ \hline
{\small A8} & {\small \  \ }$\  \ (-,+,+)$ & {\small \  \ }$\  \ -$ & $\mathbf{a}%
_{+}\mathbf{b}_{+}\mathbf{c}_{+}$ & $\frac{1}{8}$ \\ \hline
\end{tabular}%
\end{equation}%
So the total value of Ind(H$_{\mathbf{0\pi }}$) is equal to 1 and the total
contribution for the chiral lattice model is 8; which is equal to the number
of Dirac points; it is also equal 64 times the index contribution of a
fractional right (left) handed state.

\section{Conclusion and discussions}

In this paper, we have studied the anomalous properties of a family of
massive fermion models describing higher order topological matter. We have
considered chiral eight models and showed that they have 26 degrees of
freedom; six of them are basic ones and give the essence on the higher order
topological properties of these matter systems. We have computed the
explicit expressions of the gapless states for third order topological
category and showed that they are given by chiral fractional states
occupying the corners of the Figure \textbf{\ref{m}.} These fractional states%
\textbf{\ }carry, in addition to the charge $q_{_{\Gamma _{7}}}$ of $%
\mathcal{C}$, three other charges $q_{_{x}},q_{y},q_{_{z}}$ whose product is
equal to $q_{_{\Gamma _{7}}}$. This charge fractionalization aspect gives a
bridge between\textrm{\ }conventional chiral AZ matter\textrm{\ }and 3D
chiral higher order topological phases. We have also calculated the\
contribution of these gapless states to the topological index; we found that
the total value of IndH is 8; it is obtained by summing over all possible
contributions which is given by 64 times the contribution of each corner to
the index.

\section{Appendix}

In this appendix, which is organised in four parts (A, B, C, D), we give
some useful details regarding the topological chiral model studied in this
paper. First, we describe some known results on 3D HOT matter with focus on
the corner states (Appendix-A). Then, we show how the eight band model we
have studied in this work is a minimal model (Appendix-B). We take this
occasion to draw the main line to build non minimal models in 3D. After
that, we describe how to engineer the matter couplings in the real lattice
(Appendix-C).\emph{\ }We end this appendix by giving a comment on what we
called fractional (1/3 and 2/3) symmetries in the main text (Appendix-D).

\textbf{Appendix }$\mathbf{A}.$ \emph{3D HOT matter and corner states}
\newline
We begin by recalling that some facts about 3D topological AZ matter and 3D
HOT matter. Three dimensional topological matter of the AZ table has some
unified aspects; in particular: $\left( 1\right) $\ A real 3D lattice with
two periodic boundaries and one open direction. For the example of cubic
matter, one can imagine the periodic boundaries as given by the x- and y-
directions and the open one by the z- dimension. $\left( 2\right) $\ Gapped
states in the 3D bulk but gapless states on the boundary surface. In other
words, 3D topological AZ matter is characterised by gapped bulk states and
gapless surface states. \newline
Regarding the 3D HOT we studied in this paper, the D/D-1 correspondence is
relaxed to D/D-3. Seen that here D=3, the topological boundary states are
then given by 0D states or corners states. This picture corresponds to the
situation where all the three real (x,y,z) directions of the 3D material are
open. In this case, the boundary surface surrounding the cubic volume is a
non- regular surface in the sense it is made of: $\left( i\right) $\ 6 types
of 2D faces F namely: the upper face F$_{xy}^{+}$\ and the bottom face F$%
_{xy}^{-}$; the front F$_{yz}^{+}$\ and the behind F$_{yz}^{-}$; left F$%
_{zx}^{+}$\ and the right F$_{zx}^{-}$. $\left( ii\right) $ 12 line edges
given by: 4 segments in x-direction, 4 segments in y-direction and 4
segments in z-direction. $\left( iii\right) $\ 8 corners given by the
intersections of line edges (or also by intersections of three normal
faces). These 8 corners are particularly interesting for us as they host the
gapless states of HOT matter we are studying in this paper. Being gapless,
these corner states obey massless Dirac-like equation $D\psi _{corner}=0$
where $D$ the Dirac-like operator which in our study is given by eqs(47) and
(50). The solutions of this equation is studied in the main text.

\textbf{Appendix }$\mathbf{B}.$ \emph{Chiral multi-bands}\newline
The chiral eight band model which we have studied in this paper is a minimal
3D model that extends the well known SSH model with open boundary. It
describes gapless corner states of 3D HOT matter. To see the minimality
feature, it is interesting to recall a known result on Dirac-like matter in
diverse space dimensions. Because of the Gamma matrices of the Dirac-like
Hamiltonian, the number N of bands is quantized like $n\times 2^{d}$ where $%
2^{d}$\ is the number of components of each Dirac-like field $\mathbf{\psi }$
and n referring to the number of matter fields (the number of spinors). This
relation means that for $d=1,$ corresponding to SSH, the number N=2n with n
for conduction band and n for valence band. For the d=2 and the d=3
extensions of SSH, the number of bands is respectively given by $n\times
2^{2}=4n$ and $n\times 2^{3}=8n$. So, the eight band 3D model studied in
present paper is a minimal model (n=1). However, the generalisation of our
modeling to chiral 8n bands goes straightforwardly. The basic idea in the
construction consists in replacing one of the 2$\times $2 Pauli matrices by
the following generalised $2n\times 2n$ matrices. For example, by replacing $%
\rho _{x},$ $\rho _{y},$ $\rho _{z}$ by the following $2n\times 2n$ matrices
$R_{x},$ $R_{y},$ $R_{z}$.%
\begin{equation}
R_{x}=\left(
\begin{array}{cc}
0 & I_{n} \\
I_{n} & 0%
\end{array}%
\right) \quad ,\quad R_{y}=\left(
\begin{array}{cc}
0 & -iI_{n} \\
iI_{n} & 0%
\end{array}%
\right) \quad ,\quad R_{z}=\left(
\begin{array}{cc}
I_{n} & 0 \\
0 & -I_{n}%
\end{array}%
\right)   \tag{B1}
\end{equation}%
where $I_{n}$ is the $n\times n$ identity matrix.

\textbf{Appendix }$\mathbf{C}$. \emph{Couplings in real lattice}  \newline
The eight band Hamiltonian (35) is expressed in the reciprocal space. This
hamiltonian can be also expressed in the real space. The derivation of the
real interactions leading to (35) is commented below. First, we consider the
Dirac-like matter field $\mathbf{\psi }\left( \mathbf{r}_{i}\right) \equiv
\mathbf{\psi }_{\mathbf{r}_{i}}$ at site positions $\mathbf{r}_{i}$ (for
short $\mathbf{r}$) of the real 3D cubic lattice. This matter field has
eight components $\left( \psi _{1\mathbf{r}},...,\psi _{8\mathbf{r}}\right) .
$ Then, we denote the nearest neighbours to $\mathbf{r}$ by $\mathbf{r+a}_{j}
$ with the three cubic vectors as usual; that is $\mathbf{a}_{x}=ae_{x},$ $%
\mathbf{a}_{y}=ae_{y},$ $\mathbf{a}_{z}=ae_{z}.$ After that, we use the $%
\Upsilon ^{j}$ and $\Lambda ^{j}$ matrices of eq(7) to couple the nearest
neighbours $\mathbf{\psi }_{\mathbf{r}}$ and $\mathbf{\psi }_{\mathbf{r+a}%
_{j}}$ as follows%
\begin{equation}
\begin{tabular}{lllllll}
$\mathbf{\psi }_{\mathbf{r}}^{\dagger }\Upsilon ^{i}\mathbf{\psi }_{\mathbf{%
r+a}_{j}}$ & , & $\mathbf{\psi }_{\mathbf{r+a}_{j}}^{\dagger }\Upsilon ^{j}%
\mathbf{\psi }_{\mathbf{r}}$ & ; & $\mathbf{\psi }_{\mathbf{r}}^{\dagger
}\Lambda ^{j}\mathbf{\psi }_{\mathbf{r+a}_{j}}$ & , & $\mathbf{\psi }_{%
\mathbf{r+a}_{j}}^{\dagger }\Lambda ^{j}\mathbf{\psi }_{\mathbf{r}}$%
\end{tabular}
\tag{C.1}
\end{equation}%
In these couplings, the $\Upsilon ^{x}$ and $\Lambda ^{x}$ are linked with
the direction $\mathbf{a}_{x}$; the same thing is done for the other
matrices. Next, we engineer: $\left( 1\right) $ the terms in $\sin k_{j}$ by
using pseudo-real couplings like%
\begin{equation}
h_{1}=-i\sum \limits_{\mathbf{r},\mathbf{a}_{j}}t_{j}\left[ \mathbf{\psi }_{%
\mathbf{r}}^{\dagger }\Upsilon ^{i}\mathbf{\psi }_{\mathbf{r+a}_{j}}-\mathbf{%
\psi }_{\mathbf{r+a}_{j}}^{\dagger }\Upsilon ^{j}\mathbf{\psi }_{\mathbf{r}}%
\right]   \label{C.2}
\end{equation}%
and $\left( 2\right) $ the terms\ in $\cos k_{j}$ by using real coupling as
follows $h_{2}=\sum t_{j}^{\prime }(\mathbf{\psi }_{\mathbf{r}}^{\dagger
}\Lambda ^{j}\mathbf{\psi }_{\mathbf{r+a}_{j}}+\mathbf{\psi }_{\mathbf{r+a}%
_{j}}^{\dagger }\Lambda ^{j}\mathbf{\psi }_{\mathbf{r}}).$ The terms in $%
\Delta _{j}$ are given by $h_{3}=\sum \Delta _{j}\mathbf{\psi }_{\mathbf{r+a}%
_{j}}^{\dagger }\Lambda ^{j}\mathbf{\psi }_{\mathbf{r+a}_{j}}.$

\textbf{Appendix }$\mathbf{D}.$ "\emph{Fractional" symmetries}  \newline
Here we give a group theoretical description of the chiral symmetry $\Gamma
_{7}=\mathcal{C}=\mathcal{C}_{x}\mathcal{C}_{y}\mathcal{C}_{z}$ acting on
the hamiltonian as in eq(1) and on the functions $f_{x}$ and $g_{x}$ like in
(30). The action of the chiral operator $\mathcal{C}_{x}\mathcal{C}_{y}%
\mathcal{C}_{z}$ on the six components $f_{x},g_{x};f_{y},g_{y};f_{z},g_{z}$
appearing in the Hamiltonian (26) reads as follows%
\begin{equation}
\begin{tabular}{lllll}
$\mathcal{C}_{x}$ & : & $\left( f_{x},g_{x};f_{y},g_{y};f_{z},g_{z}\right) $
& $\quad \rightarrow \quad $ & $\left(
-f_{x},-g_{x};f_{y},g_{y};f_{z},g_{z}\right) $ \\
$\mathcal{C}_{y}$ & : & $\left( f_{x},g_{x};f_{y},g_{y};f_{z},g_{z}\right) $
& $\quad \rightarrow \quad $ & $\left(
f_{x},g_{x};-f_{y},-g_{y};f_{z},g_{z}\right) $ \\
$\mathcal{C}_{z}$ & : & $\left( f_{x},g_{x};f_{y},g_{y};f_{z},g_{z}\right) $
& $\quad \rightarrow \quad $ & $\left(
f_{x},g_{x};f_{y},g_{y};-f_{z},-g_{z}\right) $%
\end{tabular}
\tag{D.1}
\end{equation}%
By using group theory language, each transformation $\mathcal{C}_{i}$\textrm{%
\ }generates a symmetry group\textrm{\ }$\mathbb{Z}_{2}$ with elements as $%
\left \{ I_{id},-I_{id}\right \} $. So, the combination of the three
transformations (A1) generate the discrete group product\emph{\ }$\mathbb{Z}%
_{2}\times \mathbb{Z}_{2}\times \mathbb{Z}_{2}$. For convenience, we denote
this symmetry group like $\mathbb{Z}_{2}^{x}\times \mathbb{Z}_{2}^{y}\times
\mathbb{Z}_{2}^{z}$ where we have exhibited the x- , y- and z- directions.
The first\emph{\ }$\mathbb{Z}_{2}^{x}$ is generated by $\mathcal{C}_{x}$,
the second by $\mathcal{C}_{y}$ and the third by $\mathcal{C}_{z}$. Seen
that $\mathbb{Z}_{2}\times \mathbb{Z}_{2}\times \mathbb{Z}_{2}$ is the full
chiral symmetry, the sub-symmetry\emph{\ }$\mathbb{Z}_{2}^{x}$\textrm{\ }%
generated by $\mathcal{C}_{x}$ is then a particular sub-symmetry (1/3
symmetry for short). Similarly, the sub-summetry\textrm{\ }$\mathbb{Z}%
_{2}^{x}\times \mathbb{Z}_{2}^{y}$\textrm{\ }generated by the composition $%
\mathcal{C}_{x}\mathcal{C}_{y}$ is 2/3 chiral symmetry of the full $\mathcal{%
C}$. Notice that similar things can be also said about the other 1/3
symmetries namely $\mathbb{Z}_{2}^{y}$ and $\mathbb{Z}_{2}^{z}$ as well as
about the 2/3 symmetries\textrm{\ }$\mathbb{Z}_{2}^{x}\times \mathbb{Z}%
_{2}^{z}$ and $\mathbb{Z}_{2}^{y}\times \mathbb{Z}_{2}^{z}$\textrm{.}%
\begin{equation*}
\text{ \  \ }
\end{equation*}

\begin{acknowledgement}
Professors Lalla Btissam Drissi and El Hassan Saidi would like to
acknowledge \textquotedblright Acad\'{e}mie Hassan II des Sciences et
Techniques-Morocco\textquotedblright \ for financial support. They thank
also Felix von Oppen for stimulating discussions. L. B. Drissi acknowledges
the Alexander von Humboldt Foundation for financial support via the Georg
Forster Research Fellowship for experienced scientists (Ref 3.4 - MAR -
1202992).
\end{acknowledgement}

\end{document}